# Nanoparticle Transport in Cellular Blood Flow

Zixiang Liu[§], Yuanzheng Zhu[§], Rekha R. Rao[¶], Jonathan R. Clausen[¶], Cyrus K. Aidun[§*]

[§] George W. Woodruff School of Mechanical Engineering, Georgia Institute of Technology, Atlanta, GA 30332, USA

[¶] Sandia National Laboratories, Albuquerque, NM 87185, USA

## ABSTRACT

The biotransport of the intravascular nanoparticle (NP) is influenced by both the complex cellular flow environment and the NP characteristics. Being able to computationally simulate such intricate transport phenomenon with high efficiency is of far-reaching significance to the development of nanotherapeutics, yet challenging due to large length-scale discrepancies between NP and red blood cell (RBC) as well as the complexity of NP dynamics. Recently, a lattice-Boltzmann (LB) based multiscale simulation method has been developed to capture both NP scale and cellular level transport phenomenon at high computational efficiency. The basic components of this method include the LB treatment for the fluid phase, a spectrin-link method for RBCs, and a Langevin dynamics (LD) approach to capturing the motion of the suspended NPs. Comprehensive two-way coupling schemes are established to capture accurate interactions between each component. The accuracy and robustness of the LB-LD coupling method are demonstrated through the relaxation of a single NP with initial momentum and self-diffusion of NPs. This approach is then applied to study the migration of NPs in a micro-vessel under physiological conditions. It is shown that Brownian motion is most significant for the NP distribution in $20\,\mu m$ vessels. For $1 \sim 100\,nm$ particles, the Brownian diffusion is the dominant radial diffusive mechanism compared to the RBC-enhanced diffusion. For $\sim 500\, nm$ particles, the Brownian diffusion and RBC-enhanced diffusion are comparable drivers for the particle radial diffusion process.

*Author to whom correspondence should be addressed, cyrus.aidun@me.gatech.edu

## 1. INTRODUCTION

Analysis of whole blood flow has been demonstrated through direct numerical simulation (DNS) of the major constituents of blood including the plasma, red blood cells (RBCs) (~45%), and other cells (~0.7%) such as white blood cells (WBCs) and platelets [1-5]. Both single RBC dynamics [1, 3, 6] and rheological properties of dense suspensions of RBCs [7-10] have been computationally resolved, showing promising agreements with experimental results. Particularly, the lattice-Boltzmann (LB) method for the fluid phase coupled with the Spectrin-Link (SL) analysis of the RBC membrane as a hybrid mesoscopic method (LB-SL) has shown to be both efficient and accurate [3].

Owing to the success of DNS for whole blood, the mechanisms of migration and margination of microscale particles in blood flow have been understood to a considerable extent. In the study of hemostasis and platelet-rich thrombi formation, the mechanism of platelet margination has been investigated and shown its dependence on hemodynamics and cell properties [11-13]. For micro-sized particles of high rigidity (such as platelets, WBCs and stiffened RBCs under pathological conditions) suspended in non-dilute RBC suspensions, the propensity of the particle margination is found to be mainly driven by the RBC-enhanced diffusion in the RBC-laden region as well as the sink-like effect of the RBC-depleted layer [14-16]. Based on these complex margination mechanisms, a continuum model has been proposed to bridge the DNS capability with the patient-specific applications [17].

In contrast, analysis of the transport of NPs in cellular blood flow remains challenging due to the large length-scale discrepancy between NPs $\sim O(10\,nm)$ and RBCs $\sim O(10\,\mu m)$. Moreover, further complexities come from the intricate NP dynamics, which highly depend on the particle Brownian effect, inter-particle hydrodynamic interactions (HI) mediated through the fluid, and particle-RBC interactions. Because of the rapid development of the nanotherapeutics field, more attention has been drawn to understand the NP transport in blood vessels.

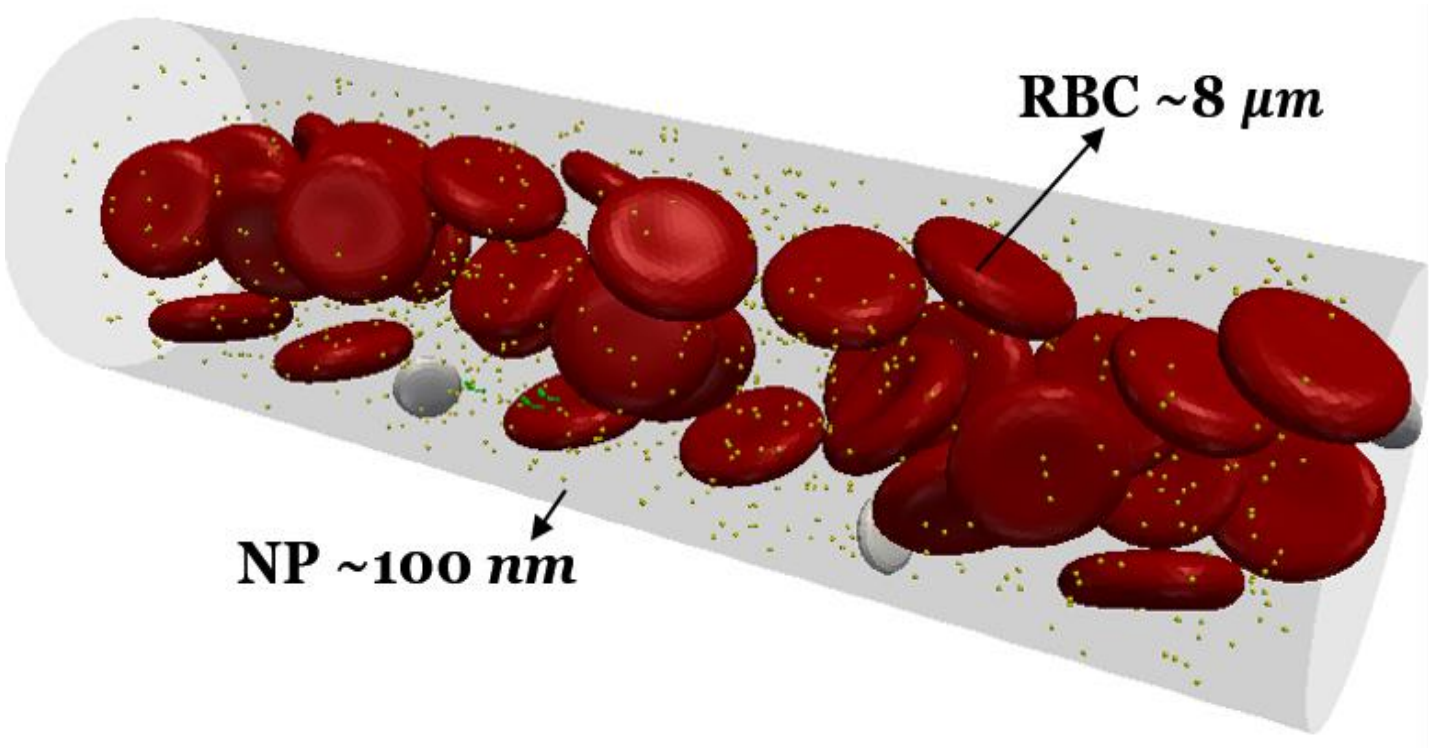

**Figure 1.** NPs (yellow dots) transport in a cellular blood vessel with various constitutive components of large length-scale discrepancies. RBC (red) is about $8\,\mu m$; platelet (white) is $2\sim3\,\mu m$; the monomers of certain proteins, such Von Willebrand factors (VWFs), can be as small as ~$60\,nm$; and NPs are generally in the range of $0\sim500\,nm$.

To study the influence of RBCs on NP transport in microcirculation, Tan at. al. [18] apply a simplified Brownian dynamics approach for NPs and an immersed finite-element (FE) method for both RBC deformation and fluid flow, showing substantial margination behavior for $100\,nm$ particles. Although the Brownian effect is included in the method, particles are only treated as passive tracers without including the effect of HI. It is shown that HI has a significant effect on the microstructure and rheological properties of colloidal/non-colloidal suspensions [19, 20]. A similar hydrodynamic approach is applied by Lee et al. [21] with all constituents (including NPs) resolved directly using FE grids. In their simulation, a dispersion factor defined as the ratio between the calculated radial diffusivity and theoretical Brownian diffusivity is introduced to quantify the severity of particle margination, which falls short of describing the actual total diffusivity of the particle. A rather insignificant NP margination is observed in their study, which is contradictory to the margination behaviors observed by Tan et al. [18]. Viewing the inefficiency of the current three-dimensional whole blood flow solvers, Tan et al. [22] apply two-dimensional simulations to obtain parametric behaviors of NP transport in cellular blood flow, which provides limited understanding on the actual three-dimensional NP migration behaviors; although some efforts have been made by Muller et al. [23] to try to connect the two-dimensional particle margination behavior with the three-dimensional counterparts.

One explanation for the above contradictory prediction of NP distribution in vessels could be overlooking the Brownian effect on the NP dynamics, given none of the studies above have provided solid verification or detailed analysis of the NP Brownian motion. However, the effect of thermal fluctuation on NPs suspended in blood is fundamentally important. For example, NP of diameter $50{\sim}100\,nm$ suspended in a $20\,\mu m$ vessel under typical wall shear rate $\dot{\gamma}_w = 500\,s^{-1}$ yields a Péclet number, $Pe = 3\pi\mu\dot{\gamma}_w d_P^3/(4k_B T)$, in the range of $0.04{\sim}0.3$, indicating the significance of NP Brownian effect. Therefore, it is paramount to correctly resolve the NP Brownian motion in order to predict the accurate NP biodistribution in cellular vessels.

Given the multiscale nature of NP transport in cellular blood flow, hybrid approaches that combine NP dynamics and mesoscale hydrodynamic approach may be the key to realize accessible three-dimensional parametric studies with large variable space. Ahlrichs et al. [24] couples a fluctuating LB method [25] with a Molecular dynamics (MD) approach for point particles through a friction term, exhibiting promising efficacy in dealing with solvent-polymer systems. Compared with the typical Brownian dynamics approach that addresses HI by dealing with an expensive mobility matrix, this hybrid LB-MD approach scales linearly with the number of particles; however, it requires an empirical rescaling of the prescribed friction coefficient to produce the theoretical Brownian diffusivity. Recently, Mynam et al. [26] showed that the empirical modification of the friction coefficient is due to extra mobility introduced by the fluctuating LB method. By removing the fluctuation in the fluid phase, a particle Brownian diffusivity is correctly determined without any artificial rescaling. Previous hybrid approaches for particle-solvent systems [24, 26-28] require sub-iterations to maintain numerical stability while solving the Langevin equation (LE) coupled with the LB method, reducing overall computational efficiency.

Here, an efficient three-dimensional multiscale LB-LD approach for the NP-solvent system coupled with the well-established LB-SL method for RBC suspension [3] is proposed to fully resolve the NP transport in cellular blood flow. Two-way coupling between the NP and fluid phase is achieved by introducing an LB forcing source term [29] to account for the momentum exchange between the LB and LD system. The framework of this multiscale computational approach is depicted in Figure 2, where all modules in different scales are included. The entire system is advanced in LB time scale without the necessity of introducing sub-time steps.

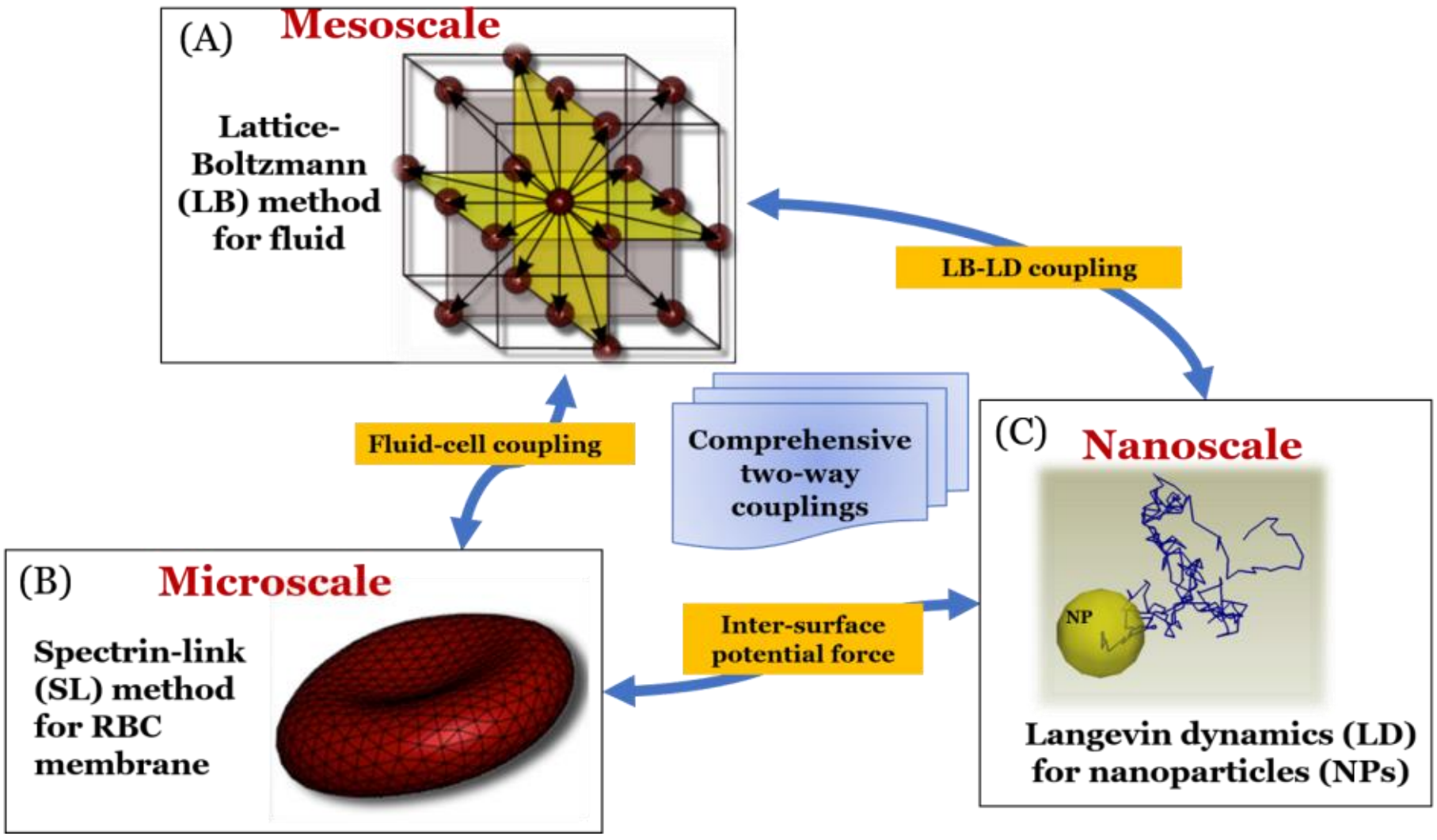

**Figure 2.** The framework of the current multiscale computational approach for simulating NP transport in cellular blood flow: (A) The lattice-Boltzmann method [2, 30, 31] for fluid phase in mesoscale; (B) the course-grained spectrin-link method [3, 7] for simulating RBC deformation and dynamics in microscale; (C) the Langevin-dynamics approach for simulating nanoscale particle dynamics and interactions. The fluid-cell coupling is based on the standard bounce-back method described in Section 2.4.1; the RBC-particle coupling is based on inter-surface potentials such as Morse potential and Lennard-Jones potential, shown in Section 2.4.3; the fluid-particle coupling is via the LB-LD coupling algorithm presented in Section 2.4.2. This LB-LD approach can be readily extended to simulate polymer-solvent systems [32, 33].

In Section 2, the basic elements of this computational approach are presented, namely the LB method for the fluid phase, the coarse-grained SL method to capture the RBC membrane dynamics, the LD approach for NPs, and the comprehensive two-way coupling schemes that bridge the entire computational framework. In Section 3, the accuracy and robustness of this approach are demonstrated through multiple benchmark cases. Then in Section 4, this approach is applied to study the NP migration in a post-capillary vessel with a physiological concentration of RBCs. In Section 5, this work is concluded with some remarks and an outlook to the future work.

## 2. METHODOLOGY

### 2.1. *Lattice-Boltzmann method*

The method used to solve for the fluid-phase is based on the three-dimensional LB method developed by Aidun et. al [2, 30, 31]. The LB method relies on propagating the fluid particles with discrete velocities, $\boldsymbol{e}_i$, resulting in the formation of a lattice space. A collision step relaxes the particle distribution function (PDF), $f_i$, towards a local equilibrium PDF, $f_i^{(eq)}$, causing a diffusion of momentum. With the collision term linearized by the single-relaxation-time Bhatnagar, Gross, and Krook (BGK) operator [34], the time evolution of the PDF takes the form of

$$f_i(\boldsymbol{r} + \Delta t_{LB}\boldsymbol{e}_i, t + \Delta t_{LB}) = f_i(\boldsymbol{r}, t) - \frac{1}{\tau}\left[f_i(\boldsymbol{r}, t) - f_i^{(eq)}(\boldsymbol{r}, t)\right] + f_i^S(\boldsymbol{r}, t), \qquad (1)$$

where $\tau$ is the normalized single relaxation time, and $f_i^S$ is a forcing source PDF based on the method of He et al [29] to account for the external force effect. This method has a pseudo-sound-speed of $c_s = \frac{\Delta r_{LB}}{\sqrt{3}\Delta t_{LB}}$, and a kinematic viscosity of $\nu_{LB} = \left(\tau - \frac{1}{2}\right) c_s^2 \Delta t_{LB}$ [35], where the LB time step, $\Delta t_{LB}$, and lattice unit distance, $\Delta r_{LB}$, are set equal to 1. At low Mach number, i.e., small $\frac{u}{c_s}$, the LB equation recovers the incompressible Navier-Stokes equation [36] with the equilibrium PDF determined by local macroscopic variables as

$$f_i^{(eq)}(\boldsymbol{r},t) = \omega_i \rho_{LB} \left[ 1 + \frac{1}{c_s^2}(\boldsymbol{e}_i \cdot \boldsymbol{u}) + \frac{1}{2c_s^4}(\boldsymbol{e}_i \cdot \boldsymbol{u})^2 - \frac{1}{2c_s^2}(\boldsymbol{u} \cdot \boldsymbol{u}) \right], \quad (2)$$

where $\rho_{LB}$ and $\boldsymbol{u}$ are the macroscopic fluid density and velocity given by moments of the equilibrium distribution functions, i.e., $\rho_{LB} = \sum_{i=1}^{\mathbb{N}} f_i^{(eq)}$ and $\boldsymbol{u} = \frac{1}{\rho_{LB}} \sum_{i=1}^{\mathbb{N}} f_i^{(eq)} \boldsymbol{e}_i$. The lattice weights, $\omega_i$, are determined by the LB stencil in use. For the D3Q19 stencil used in this study, $\mathbb{N}$ is equal to $19$, and $\omega_i$ is $1/3$, $1/18$, and $1/36$ for the rest, nondiagonal, and diagonal directions, respectively [2]. The LB method is extensively validated [31, 37, 38] and proved to be suitable for DNS of dense suspension of particles and capsules [2, 4].

### 2.2. *Spectrin-link method*

The SL method models the RBC membrane as a surface consisting of a triangular network. A high-resolution study [39] using the SL method with spectrin length on the order of proteins in actual RBCs requires more than 25,000 vertices per RBC. This level of resolution is not suitable for simulations of blood flow with realistic hematocrit containing many RBCs. The use of a coarse-grained SL approach [1, 40] coupled with the LB method allows for accurate analysis of cellular blood flow at physiological concentrations [3]. The LB-SL method has been extensively validated with experimental results and proved to be an effective tool to capture both single RBC dynamics and rheology of dense suspensions of RBCs [2, 3, 7].

In the SL model, the RBC membrane is modeled as a triangulated network, consisting of $N_v$ vertices. The vertices of each surface located at $\boldsymbol{x}_n$ $(n \in 1, \ldots, N_v)$, connected by $N_s$ spring links with lengths of $l_i$ $(i \in 1, \ldots, N_s)$ to form $N_t$ triangles with areas of $A_k$ $(k \in 1, \ldots, N_t)$. The Helmholtz free energy of the network system $E(\boldsymbol{x}_n)$, including plane components and constraint potentials to conserve membrane area and cell volume, is given as

$$E(\boldsymbol{x}_n) = E_{in-plane} + E_{bending} + E_{volume} + E_{area}. \tag{3}$$

The in-plane free energy, $E_{in-plane}$, consisting of the contributions of elastic energy stored in spectrin proteins and hydrostatic elastic energy stored in the membrane, reads

$$E_{in-plane} = \sum_{i \in 1,\ldots,N_s} U_{WLC}(l_i) + \sum_{k \in 1,\ldots,N_t} \frac{C_q}{A_k^q}. \tag{4}$$

Here, $U_{WLC}$ is the elastic energy modeled with the compressional form of worm-like chain (WLC) potential [41, 42]

$$U_{WLC}(l_i) = \frac{k_B T l_{\max}}{4p} \frac{3\hat{l}_i^2 - 2\hat{l}_i^3}{1 - \hat{l}_i}, \tag{5}$$

where $k_B$ is the Boltzmann constant, $T$ is absolute temperature, $p$ is the persistence length, and $\hat{l}_i = \frac{l_i}{l_{max}} \in [0,1]$ is the fractional link extension, where $l_{\max}$ is the maximum length of the spectrin-link. The second term in equation (4), introduced to balance the compressional effect of WLC potential, represents a repulsive potential due to the hydrostatic energy stored in the membrane patches. The constant $C_q$ is determined through the virial theorem by setting the Cauchy stress to zero [1, 43]

$$C_q = \frac{\sqrt{3} A_{l_0}^{q+1} k_B T \left(4\hat{l}_0^2 - 9\hat{l}_0 + 6\right)}{4pq l_{max} \left(1 - \hat{l}_0\right)^2}, \tag{6}$$

where $\hat{l}_0 = \frac{l_0}{l_{max}}$, $l_0$ is the average length of the links at equilibrium, and $A_{l_0} = \frac{\sqrt{3}l_0^2}{4}$. The current study selects $q = 1$, implying the membrane area modulus is always twice of its shear modulus [39]. The bending energy is defined as

$$E_{bending} = \sum_{j\in 1,\dots,N_s} \hat{k}\,[1 - \cos(\theta_j - \theta_0)], \tag{7}$$

where $\hat{k}$ is the discrete bending coefficient, $\theta_j$ is the instantaneous angle between adjacent triangles sharing the link $j$, and $\theta_0$ is the spontaneous angle. The discrete bending modulus can be related to the average bending modulus $k$ by $\hat{k} = \frac{2k}{\sqrt{3}}$ [43]. The volume conservation constraint is introduced to ensure the volume of the RBC enclosed by the triangulated membrane remains constant

$$E_{volume} = \frac{k_v\left(\Omega - \Omega_0^d\right)^2}{2\Omega_0^d}, \tag{8}$$

where $\Omega$ is the instantaneous volume of the RBC and $\Omega_0^d$ is the desired volume of the RBC. Similarly, the constraint on area serves to conserve the surface area of the RBC and takes the form of

$$E_{area} = \frac{k_A\left(A_t - A_{t,0}^d\right)^2}{2A_{t,0}^d}, \tag{9}$$

where $A_t$ is the total instantaneous area of the RBC membrane, $A_t = \sum_{i\in 1,\dots,N_t} A_i$, and $A_{t,0}$ is the desired membrane area.

The forces due to the SL method via Helmholtz free energy contributions are then determined by

$$\mathbf{f}_n^{SL} = -\frac{\partial E(\boldsymbol{x}_n)}{\partial \boldsymbol{x}_n}. \tag{10}$$

Each of the vertices that combine to form the triangulated RBC membrane surface advances per the Newton's equation of motion given by,

$$\frac{d\boldsymbol{x}_n}{dt} = \boldsymbol{v}_n\,, \tag{11}$$

$$M\frac{d\boldsymbol{v}_n}{dt} = \mathbf{f}_n^{SL} + \mathbf{f}_n^{LB} + \mathbf{f}_n^{PP}, \tag{12}$$

where $\boldsymbol{v}_n$ is the velocity of vertex $n$ and $M$ is taken to be a fictitious mass at each point chosen to yield the same membrane mass [8]. The external forces $\mathbf{f}_n^{LB}$ are forces on the vertex due to the fluid-solid interaction. $\mathbf{f}_n^{PP}$ is the force due to particle-particle (PP) interactions [44]. The locations of the vertices are updated via Newton's equations of motion using a first-order accurate forward-difference Euler scheme.

### 2.3. *Langevin dynamics*

Particles suspended in fluid are subjected to the impacts of the randomly fast-moving liquid molecules owing to the thermal fluctuation. For particles at nanoscale or submicron-scale, such instantaneously fluctuating momentum transferred from the solvent molecules spurs the particle to yield irregular movements, which is known as Brownian motion. The dynamics of such Brownian particles can be described via the LE

$$m_i\frac{d\boldsymbol{u}_p^i}{dt} = \boldsymbol{C}_p^i + \boldsymbol{F}_p^i + \boldsymbol{S}_p^i\,, \tag{13}$$

where $m_i$ is the mass of a particle $i$. Given the particle initial position, $\boldsymbol{r}_{p,0}^i$, the displacement of the particle can be readily updated through the temporal integration of the particle velocity, $\boldsymbol{r}_p^i = \boldsymbol{r}_p^i + \int \boldsymbol{u}_p^i dt$. The particles are assumed to be neutrally buoyant given the dominant fluid viscous effect. The right-hand-side (RHS) of Equation (13) typically consists of three driving forces for the particle dynamic system.

The conservative force, $\boldsymbol{C}_p^i$, specifies the interparticle or particle-surface interaction force that is determined by differentiating the total potential energy $U_{total}^i$ with respect to the particle position, $\boldsymbol{r}_p^i$,

$$\boldsymbol{C}_p^i = -\frac{dU_{total}^i}{d\boldsymbol{r}_p^i}, \tag{14}$$

The components of $U_{total}^i$ are elaborated in Section 2.4.3 and Section 2.4.4.

The friction force $\boldsymbol{F}_p^i$ due to the relative motion of the particle with respect to the local viscous fluid is assumed to follow the Stokes formula [24],

$$\boldsymbol{F}_p^i = -\zeta\left[\boldsymbol{u}_p^i(t) - \boldsymbol{u}\left(\boldsymbol{r}_p^i, t\right)\right], \tag{15}$$

where $\boldsymbol{u}_p^i$ denotes the particle velocity and $\boldsymbol{u}\left(\boldsymbol{r}_p^i, t\right)$ is the fluid velocity at the particle site. Since the particle moves continuously across the discrete lattice, $\boldsymbol{u}\left(\boldsymbol{r}_p^i, t\right)$ needs to be interpolated from the neighboring LB nodes as illustrated in the Section 2.4.2. Given the sub-grid scale and low-Reynolds number features of the Brownian particle, the friction coefficient, $\zeta$, is assumed to follow the lower-order Stokes' drag law

$$\zeta = 3\pi\mu d_p k_v, \tag{16}$$

where $\mu$ is the dynamic viscosity of the blood plasma; the particle shape factor, $k_v$, is set to unity for spherical particles.

The stochastic force term, $\boldsymbol{S}_p^i$, accounts for the thermal fluctuation of the solvent and serves as the source of the Brownian effect of the particle. Through the equipartition principle and the integration of the LE with respect to time [45], the stochastic force can be related to friction, demonstrating a balance between the random thermal fluctuation and the frictional dissipation, i.e., the fluctuation dissipation theorem (FDT) [46]. Specifically, the stochastic force $\boldsymbol{S}_p^i$ yields the following statistical properties

$$\langle S_{p,\alpha}^i(t) \rangle = 0, \tag{17}$$

$$\langle S_{p,\alpha}^i(t) S_{p,\beta}^j(t') \rangle = 2k_B T \zeta \delta_{ij} \delta_{\alpha\beta} \delta(t - t'). \tag{18}$$

Here, $\alpha, \beta \in \{x, y, z\}$, $i$ and $j$ run through all the particle indices, $\delta_{ij}$ and $\delta_{\alpha\beta}$ are Kronecker deltas, $\delta(t - t')$ is the Dirac-delta function, $k_B$ is the Boltzmann constant, $T$ is the temperature of the fluid bath, and $\langle\ \rangle$ denotes the average over the ensemble of realizations of the random variables. Equation set (17) and (18) statistically states that the Cartesian components of $\boldsymbol{S}_p^i$ exhibit Gaussian distributions with zero means.

To avoid excessive computational expense when solving the LE, it is ideal to temporally discretize the LE with $\Delta t_{LB}$. This can be achieved by conditionally neglecting particle inertia in the LE to yield correct particle dynamics as well as maintaining numerical stability. Two various time scales need to be considered, namely, the Brownian relaxation time scale, $\tau_r = \frac{m}{\zeta}$, over which the particle velocity decays to the algebraic long-time tail regime, and the LB time scale, $\tau_{LB} = \frac{L^2 \nu_{LB}}{L_{LB}^2 \nu} \Delta t_{LB}$, which determines the physical time interval corresponding to each LB time step. The LB grid resolution, $\frac{L}{L_{LB}}$, is set to 333 $nm/lu$, which is fine enough to capture the accurate RBC deformation and dynamics as shown in previous studies [3, 7]; the kinematic viscosity ratio, $\frac{\nu}{\nu_{LB}}$, is determined by the fluid kinematic viscosity, $\nu$, and the LB relaxation time, $\tau$, in use. Specifically, when $\tau_{LB} > \tau_r$ , to avoid numerical instability without introducing sub-time steps, the over-damped LE can be solved, assuming particle velocity has fully relaxed (with zero inertial effect) after each time step; when $\tau_{LB} \leq \tau_r$, since physically the particle has not relaxed to its terminal velocity after each $\Delta t_{LB}$, the under-damped LE can be advanced with $\Delta t_{LB}$ to include the particle short-time inertial effect while maintaining numerical stability [47]. In accordance with the temporal discretization of the LB and SL governing equations, forward-differencing Euler method with first-order accuracy is employed to advance the LE in time. Therefore, the Brownian particle velocity and displacement can be advanced, based on the time scales, by

$$\boldsymbol{u}_p(t + \Delta t_{LB}) = \begin{cases} \boldsymbol{u}(\boldsymbol{r}_p, t) + \frac{1}{\zeta}\left[\boldsymbol{C}_p(t) + \boldsymbol{S}_p(t)\right], & (\tau_{LB} > \tau_r) \\ \boldsymbol{u}_p(t) + \frac{\Delta t_{LB}}{m}\left\{\boldsymbol{C}_p(t) + \boldsymbol{S}_p(t) - \zeta\left[\boldsymbol{u}_p(\boldsymbol{t}) - \boldsymbol{u}(\boldsymbol{r}_p, t)\right]\right\}, & (\tau_{LB} \leq \tau_r) \end{cases} \tag{19}$$

$$\boldsymbol{r}_p(t + \Delta t_{LB}) = \boldsymbol{r}_p(t) + \Delta t_{LB}\boldsymbol{u}_p(t + \Delta t_{LB}). \tag{20}$$

By conditionally solving the LE, the particle short-time inertial effect is handled selectively without the need to compromise to sub-time steps as utilized in previous studies [24, 26-28]. The switch between two forms of the discretized LE (19) is based on the time scale difference. The Gaussian distribution associated with the stochastic force $\boldsymbol{S}_p(t)$ is realized via a random number generator based on the Box-Muller transformation [48].

## 2.4. *Coupling schemes*

### 2.4.1. *RBC-fluid coupling*

The coupling between the fluid and RBCs is established via a moving boundary bounce-back method [31]. In this method, the momentum transfer at the fluid-solid interface is accounted for by applying the standard bounce-back (SBB) scheme along lattice links that cross solid surfaces. Using the bounce-back method, the no-slip condition is enforced by adjusting the distributions of fluid nodes at the end points of a link in the $i$ direction via

$$f_{i'}(\boldsymbol{r}, t+1) = f_i(\boldsymbol{r}, t^+) - 6\rho\omega_i \boldsymbol{u}_b \cdot \boldsymbol{e}_i, \tag{21}$$

where $i'$ is the direction opposite to $i$, $f_i(\boldsymbol{r}, t^+)$ is the post-collision distribution, and $\boldsymbol{u}_b$ is the solid velocity at the intersection point with the link. The fluid force exerted on the RBC surface vertex is thus determined by

$$\mathbf{f}_n^{LB}\left(\boldsymbol{r} + \frac{1}{2}\boldsymbol{e}_i, t\right) = 2\boldsymbol{e}_i[f_i(\boldsymbol{r}, t^+) + 3\rho\omega_i \boldsymbol{u}_b \cdot \boldsymbol{e}_{i'}]. \tag{22}$$

### 2.4.2. *NP-fluid coupling*

The LD for NP and the LB fluid are coupled in a two-way fashion through exchange of fluid forces exerted on the NP [27]. To update the NP dynamics through the LE, the fluid force exerted on the particle, $\boldsymbol{F}_p^f$, can be decomposed into a frictional force, $\boldsymbol{F}_p$, and a stochastic force, $\boldsymbol{S}_p$, such as

$$\boldsymbol{F}_p^f = \boldsymbol{F}_p + \boldsymbol{S}_p = -\zeta\left[\boldsymbol{u}_p(t) - \boldsymbol{u}(\boldsymbol{r}_p, t)\right] + \boldsymbol{S}_p(t). \tag{23}$$

Since each NP is treated as a sub-grid point particle and moves continuously in the lattice domain, the fluid velocity in the location of the particle site, $\boldsymbol{u}(\boldsymbol{r}_p, t)$, needs to be interpolated from the discrete velocities of the surrounding lattice nodes, as shown in Figure 3.

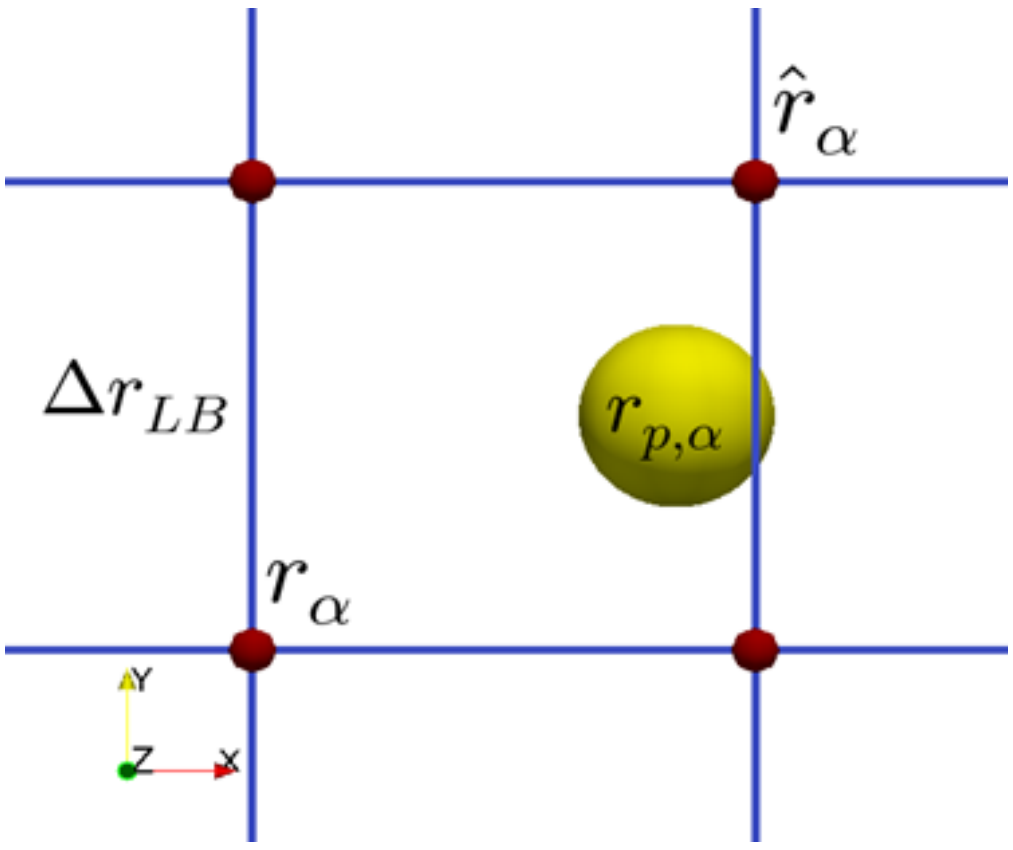

**Figure 3.** 2-D representation of a NP (yellow) located at a position, $\boldsymbol{r}_p$, in a lattice cell. Lattice node, $\hat{\boldsymbol{r}}$, denotes the diagonal node with respect to the node, $\boldsymbol{r}$, and $\alpha$ denotes the Cartesian component of each node coordinate.

In the current approach, two options of distance-based interpolation stencils, namely, the trilinear stencil [27] and Peskin's stencil [49] have been implemented to gain the flexibility of switching between schemes of different accuracy. For the trilinear stencil, a weight function can be introduced as

$$w(\boldsymbol{r}, \boldsymbol{r}_p) = \prod_{\alpha \in \{x,y,z\}} \frac{|\hat{r}_\alpha - r_{p,\alpha}|}{\Delta r_{LB}}, \tag{24}$$

which is first-order accurate and associated with at most 8 lattice nodes of the cell containing the particle. Peskin's stencil introduces a weight function by

$$w(\boldsymbol{r},\, \boldsymbol{r_p}) = \prod_{\alpha\in\{x,y,z\}} \frac{\left\{1+\cos\left[\frac{\pi(r_\alpha - r_{p,\alpha})}{2\Delta r_{LB}}\right]\right\}}{4\Delta r_{LB}}, \tag{25}$$

which can yield second-order accuracy. Peskin's stencil has also been successfully applied to the external boundary force (EBF) method [50], which shows better accuracy and efficiency compared to the SBB method in resolving fluid-solid interactions. For simplicity, the trilinear stencil will be used in the following studies. After $w(\boldsymbol{r},\, \boldsymbol{r_p})$ is calculated, $\boldsymbol{u}(\boldsymbol{r}_p, t)$ can be obtained as

$$\boldsymbol{u}(\boldsymbol{r_p}, t) = \sum_{\boldsymbol{r}\in N_c} w(\boldsymbol{r},\, \boldsymbol{r_p})\boldsymbol{u}(\boldsymbol{r}, t), \tag{26}$$

where $N_c$ denotes the group of nodes on the cell occupied by the particle.

To enforce momentum conservation of the particle-fluid system and capture the particle-fluid interaction, the reactionary impulse due to $\boldsymbol{F}_p^f$ acting on the particle needs to be assigned back to the fluid during each LB time step. By applying the same weight function, the reactionary impulse density [31] can be distributed back to the surrounding lattice nodes as

$$\boldsymbol{J}(\boldsymbol{r}, \boldsymbol{r_p}) = -w(\boldsymbol{r}, \boldsymbol{r}_p)\frac{\boldsymbol{F}_p^f \Delta t_{LB}}{\Delta r_{LB}^3}, \tag{27}$$

where $\boldsymbol{J}(\boldsymbol{r}, \boldsymbol{r}_p)$ is the impulse density to be assigned to LB node, $\boldsymbol{r}$, due to particle-fluid interaction at the particle position, $\boldsymbol{r_p}$, during each LB time step. A local forcing source distribution term, $f_i^S(\boldsymbol{r}, t)$, based on the method of He et al. [29] is then calculated as

$$f_i^S(\boldsymbol{r},t) = \frac{\omega_i \boldsymbol{J}(\boldsymbol{r},\boldsymbol{r}_p)\cdot \boldsymbol{e}_i}{c_s^2}. \tag{28}$$

Instead of modifying the local equilibrium distribution functions as shown in previous studies [24, 26, 27], the current approach, similar to the EBF method, modifies the regular LB evolution equations into Equation (1) through the addition of a LB forcing source term, $f_i^S(\boldsymbol{r},t)$, which is shown to approximate the Navier-Stokes equation in the physical timescale [51]. The validity of this coupling approach is demonstrated below in Section 3.

### 2.4.3. *NP-cell interaction*

The NP-cell interaction is established by following the method of resolving the RBC-RBC interactions through Morse potentials [10]. The Morse potential curves can be easily adjusted to match the measured inter-surface potential energy [52] as shown in Figure 4. Due to lack of statistics for actual NP-cell interaction energy, this study employs the measured RBC-RBC interaction energy [52] for the NP-cell interactions. The Morse potential function is given as

$$U_M(r_s) = D_e\left[e^{2\beta(r_s-r_e)} - 2e^{-\beta(r_s-r_e)}\right], \qquad (r_s \le r_e) \tag{29}$$

where $D_e$ is the Morse potential-well depth, $r_e$ is the equilibrium bond distance, $\beta$ is a scaling factor, and $r_s$ is the normal distance between NP and cell membrane surface or cell-cell normal surface distance. Equation (29) is truncated at the equilibrium distance to preserve repulsive behavior. In current method, $D_e = 10^7 k_B T$, $\beta = 2\ \mu m^{-1}$, and $r_e = 10\ nm$. A distance-based search algorithm is implemented to efficiently locate the nearest RBC triangulation surface with respect to each NP. The interaction between NP and vessel wall is handled similarly, except the NP-cell force is also assigned to update the RBC dynamics through $\boldsymbol{f}_n^{PP}$.

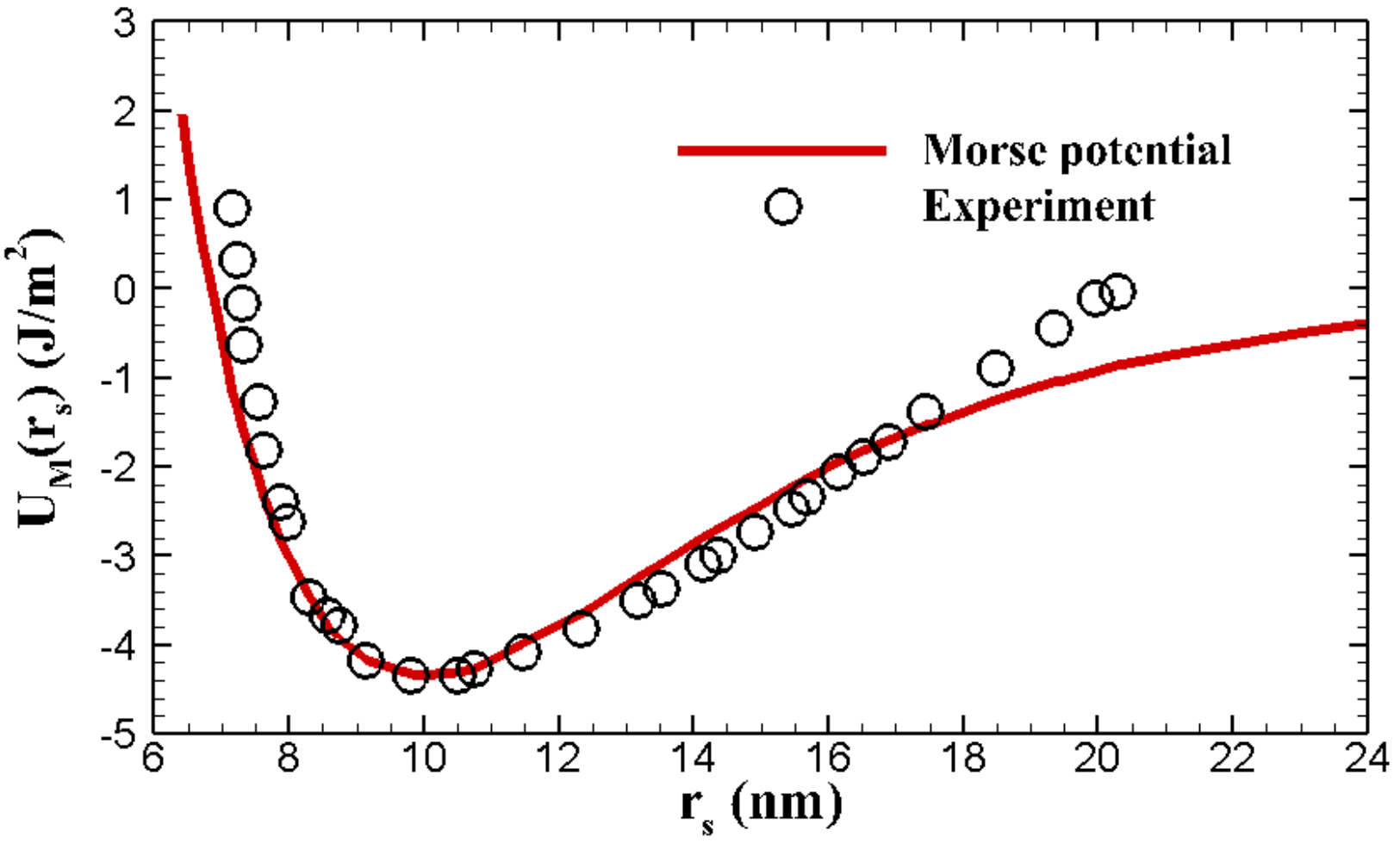

**Figure 4.** Measured cell-cell interaction energy (circle) and the fitted Morse potential function (curve) plotted against the cell-cell separation. The data is adopted from previous work with permission [53].

### 2.4.4. *NP-NP interaction*

To account for the near-field interparticle Van der Waals force and the particle volume-exclusion effect, the standard Lennard-Jones (LJ) potential, given by

$$U_{LJ}(r_{ij}) = 4\epsilon\left[\left(\frac{\sigma}{r_{ij}}\right)^{12} - \left(\frac{\sigma}{r_{ij}}\right)^{6}\right], \tag{30}$$

is implemented, where $\epsilon$ is the LJ potential-well depth chosen to be $2k_BT$, $\sigma$ is the zero-potential distance selected to be $2.5d_p$, and $r_{ij}$ is the interparticle separation. Equation (30) is truncated at an interparticle distance of $r_{ij} = 3\sigma$ to save simulation expense without losing both the attractive and repulsive effects.

All potential functions are superimposed to obtain the total potential, $U_{total}^{i}$, associated with each NP of index $i$, such as

$$U_{total}^{i} = U_{LJ}^{i} + U_{M}^{i} + U_{other}^{i}, \tag{31}$$

where $U_{other}^{i}$ refers to all the other possible potentials, such as spring potential to account for the biochemical bond effect between monomers, and Coulomb potential for the particle surface charge effect. $U_{total}^{i}$ is employed in Equation (14) to obtain the conservative force for the LD of NP. The sub-grid modeling for cell-cell short-distance interactions is discussed in detail by MacMeccan et al. [8] and Clausen et al. [44].

## 3. METHOD VERIFICATION

Since the SL method for the RBC dynamics coupled with the LB method has been extensively validated in previous studies [3, 7], Section 3 aims to address the validity of the LD approach as well as its coupling with the LB method.

### 3.1. *Velocity relaxation of a single particle*

The validity of the fluid-particle coupling is first demonstrated by analyzing the velocity relaxation process of a single particle with an initial momentum in a quiescent fluid. For all cases in this section, the stochastic force in the LE is set to zero to show deterministic response of the coupled fluid-particle system. For particles of finite mass, parameters were chosen such that $\tau_{LB} \leq \tau_r$; hence, the underdamped LE is solved to capture the particle short-time dynamic response. Since cases with $\tau_{LB} > \tau_r$ is always valid for particles of zero mass, the over-damped LE is solved accordingly. In this section, all parameters are discussed in lattice units ($lu$) without loss of generality.

First, a single particle of mass $m_1 = 29.3$ with initial velocity $u_p(0) = 0.01$ in the $x$ direction is released in a quiescent fluid. A friction coefficient $\zeta = 0.48$ is prescribed to dissipate the kinetic energy of the particle. Periodic boundary conditions are enforced on fluid domain boundaries. Three sizes of computational domains ($50^3$, $70^3$, and $100^3$) are considered to show the domain-size dependency.

Figure 5 depicts the instantaneous streamlines induced by the point particle translation in the $100^3$ fluid domain. At $t = 100$, a starting recirculation flow structure near the particle is observed, qualitatively matching the flow pattern reported by Alder et al. [54]. As the particle further slows down at $t = 300$, the radius of curvature of the instantaneous streamlines increases. When $t > 1000$, as indicated in Figure 5C, the radius of curvature reaches infinity and streamlines tend to be parallel to each other. The bulk fluid domain translates closely with the particle, reflecting a gradually vanishing momentum exchange between the particle and fluid phase.

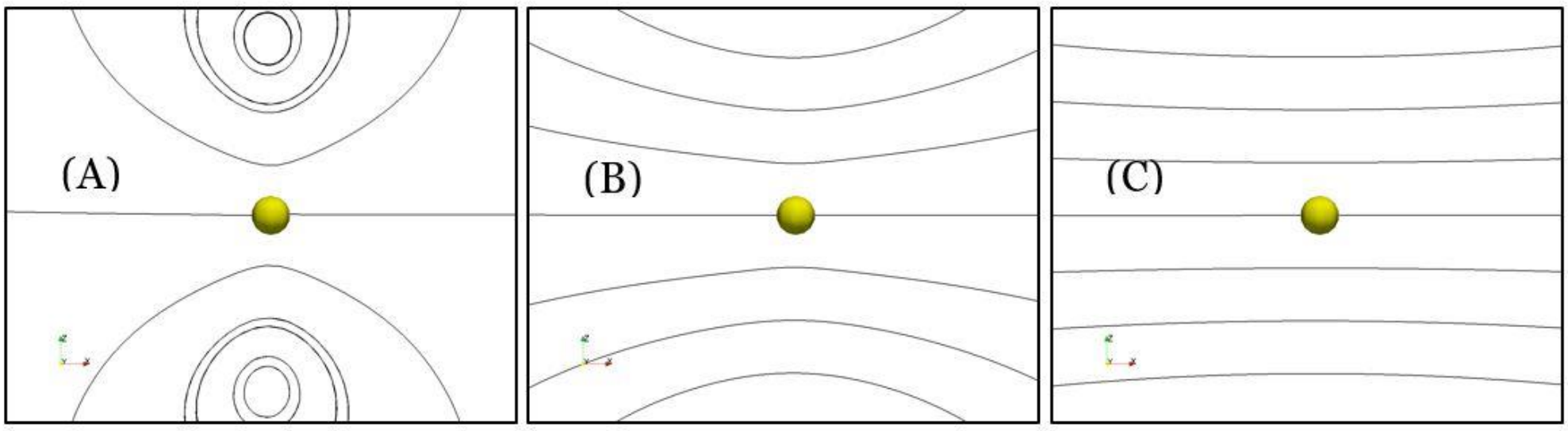

**Figure 5.** The instantaneous flow field induced by a point particle at time (A) $t = 100$, (B) $t = 300$, and (C) $t = 1000$, simulated with $m_1 = 29.3$, $\zeta = 0.48$ and $u_\mathrm{p}(0) = 0.01$ in a $100^3$ fluid domain.

In Figure 6, the time evolution of the normalized particle velocity is quantitatively compared with the well-established asymptotic behaviors. In the short-time regime, i.e., $t < 100$, the velocity of the particle is shown to decay exponentially for all three domain sizes.

This matches well with the analytical short-time asymptotic behavior, $\exp\left(-\frac{\zeta t}{m_1}\right)$. In the long-time regime, Alder and Wainwright [54] discover that the particle velocity decays by the power law, $t^{-3/2}$, as time asymptotically goes to infinity. This so-called long-time tail algebraic decay behavior is due to the hydrodynamic interactions between the NP and fluid phase. In the current study, similar long-time asymptotic behavior is observed as the domain size increases to $100^3$ at $600 < t < 3000$. This directly indicates that the hydrodynamics and the particle dynamics are correctly coupled via the LB-LD coupling algorithm as presented in Section 2.4.2. All three cases eventually reach certain finite terminal particle velocities, which is attributed to the confinement effect. The terminal velocity can be estimated based on the momentum conservation of the coupled particle-fluid system. As the computational domain expands, a more pronounced $t^{-3/2}$ behavior is observed. Therefore, to minimize the finite-domain effect with acceptable computational expense, the $100^3$ domain is chosen for the following studies.

Figure 7 presents the velocity relaxation for particles of different inertial effects. The particle inertia is either decreased by reducing the particle mass as $m_2 = 0.1m_1$ or neglected by solving the over-damped LE. As expected, the lighter particle of mass $m_2$ exhibits a faster decay with the long-time tail occurring about ten times faster than the heavier particle of mass $m_1$. For the particle with no inertia, the particle velocity directly relaxes to the long-time tail without yielding the exponential decay behavior. The fluctuations in the relaxation curves for the low/no inertia cases is due to larger discretization error in time, which can be eliminated using smaller time steps as suggested in other studies [24, 26]. For all cases of different inertial effects, the corresponding short-time or long-time asymptotic behaviors are well resolved. Particularly, the capture of the long-time scaling law, $t^{-3/2}$, suggests the far-field interparticle HI are accurately resolved as also demonstrated in other studies [24, 27, 55].

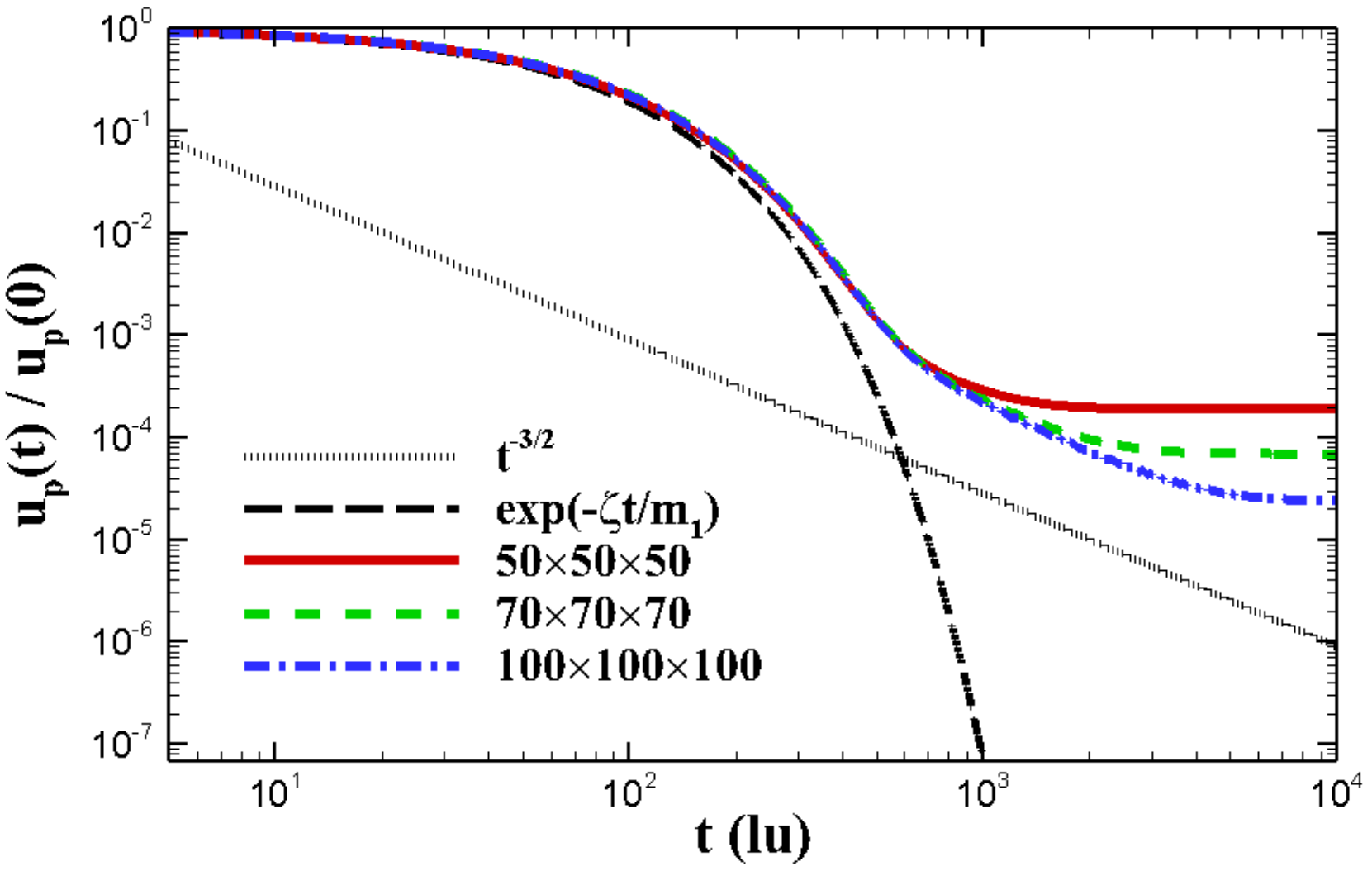

**Figure 6.** Relaxation of the normalized particle velocity, $u_p(t)/u_p(0)$, for a point particle with an initial disturbance in domains of different sizes $(50^3, 70^3\ and\ 100^3)$. The simulation was performed via under-damped LE at $m_1 = 29.3$, $\zeta = 0.48$ and $u_p(0) = 0.01$ without stochastic forces.

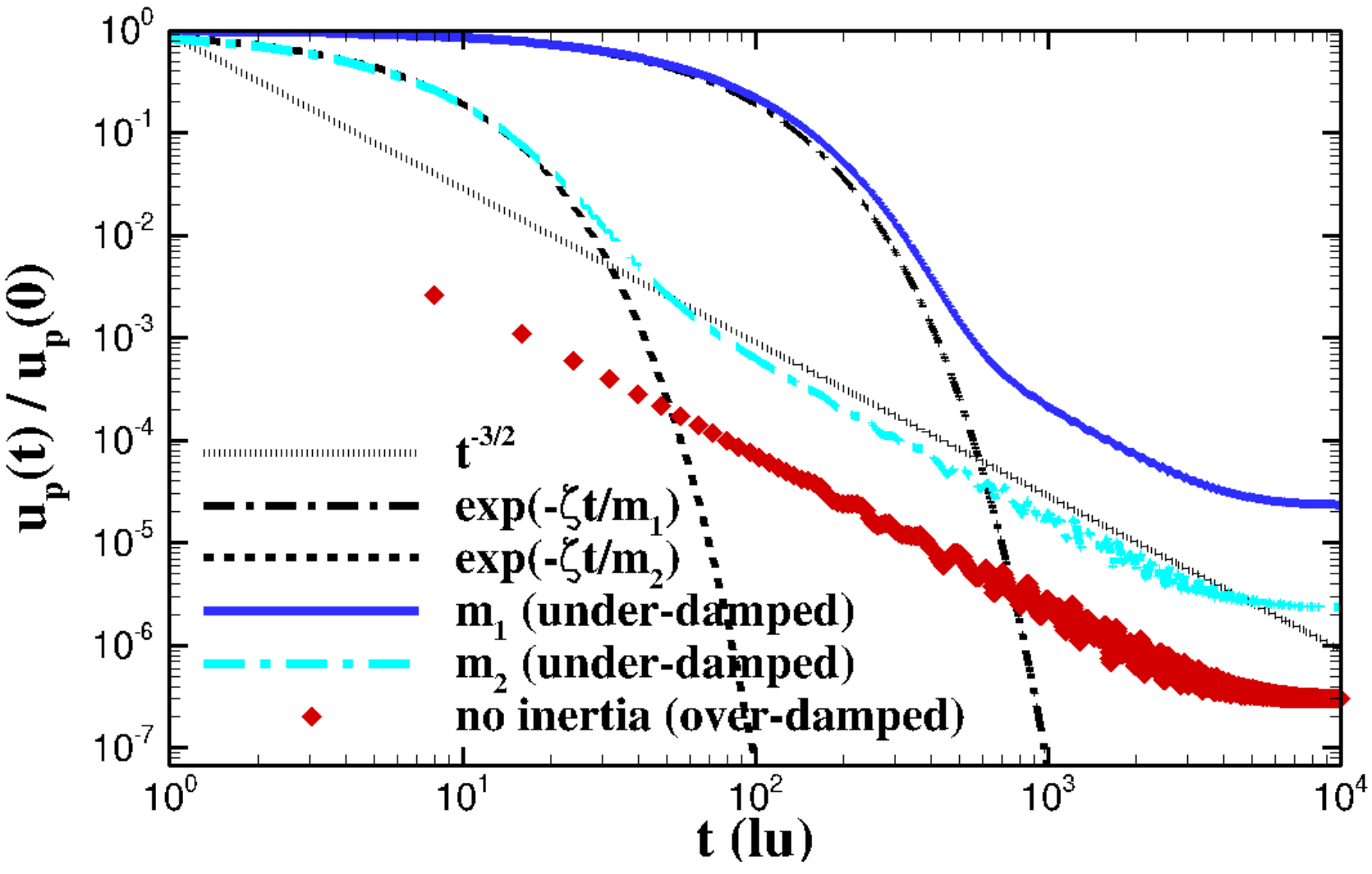

**Figure 7.** Velocity relaxation of an initially disturbed particle in the fluid, simulated with different particle inertial effects. Particles of finite mass $(m_2 = 0.1m_1)$ is simulated via under-damped LE, while over-damped LE is for the particle of zero inertia, all simulated at $\zeta = 0.48$ and $u_p(0) = 0.01$ without stochastic forces.

### 3.2. *Self-diffusion of nanoparticles*

To capture the accurate diffusive behavior of NPs due to thermal fluctuation, the long-time particle diffusivity needs to be verified to yield the correct Brownian diffusivity. For an isolated Brownian particle suspended in a quiescent fluid, the particle diffusivity is given by Einstein [56] as

$$D_{theo} = \frac{k_B T}{\zeta}, \tag{32}$$

which relates the particle diffusive behavior to the dissipative property of the particle-fluid system. Combining Equation (16) with Equation (32) produces the so-called Stokes-Einstein relation. The particle diffusivity can also be calculated through measuring the mean squared displacement (MSD) of the self-diffusive particle via

$$D_{sim} = \frac{MSD_t}{6\Delta t}\bigg|_{\Delta t \gg \tau_r}, \tag{33}$$

where $MSD_t = \langle [\boldsymbol{r}_p(t + \Delta t) - \boldsymbol{r}_p(t)]^2 \rangle_t$ denotes the time-averaged MSD measured based on the particle trajectory. Assuming the particle-fluid system satisfies the ergodic hypothesis [57], the time-averaged MSD should identically match the ensemble MSD. The sampling time interval, $\Delta t$, needs to be much greater than the particle relaxation time, $\tau_r$, to yield accurate long-time diffusivity [55].

An isolated particle of diameter $d_p = 100\,nm$ is first considered to be neutrally suspended in a liquid of dynamic viscosity $\mu = 1.2\,cP$ at room temperature $T = 298\,K$. Since the parameters yield $\tau_{LB} > \tau_r$, the over-damped LE is employed to demonstrate the long-time diffusive behavior. $10^6$ LB time steps are performed for each run to obtain converged $MSD_t$ with enough time samples. The current simulation selects $\tau = 1$, which leads to a temporal resolution of $0.0147\,\mu sec$ per LB time step. This setup also satisfies $\Delta t \gg \tau_r$ when more than ten LB time steps are used as the sampling interval.

By performing independent runs with a series of sampling time interval $\Delta t = 0.147 \sim 14.7\,\mu sec$, the corresponding rescaled MSDs, $MSD_t/6$, are evaluated and plotted in Figure 8. The solid line indicates the theoretical $MSD_t/6$ values evaluated by the Stokes-Einstein relation, which agrees well with the simulated MSD data points. Similarly, by choosing a fixed sampling time interval $\Delta t = 14.7\,\mu sec$, the particle instantaneous diffusivity $D_{sim}$ asymptotically approaches the theoretical diffusivity $D_{theo}$, as shown in Figure 9. The initial oscillation of the diffusivity curve and the slight overprediction of the long-time asymptotic diffusivity are due to the insufficiency of sampling. The snapshots of particle trajectory at different instance are also depicted in Figure 10, which is similar to the typical Brownian particle trajectory pattern reported by other studies [58].

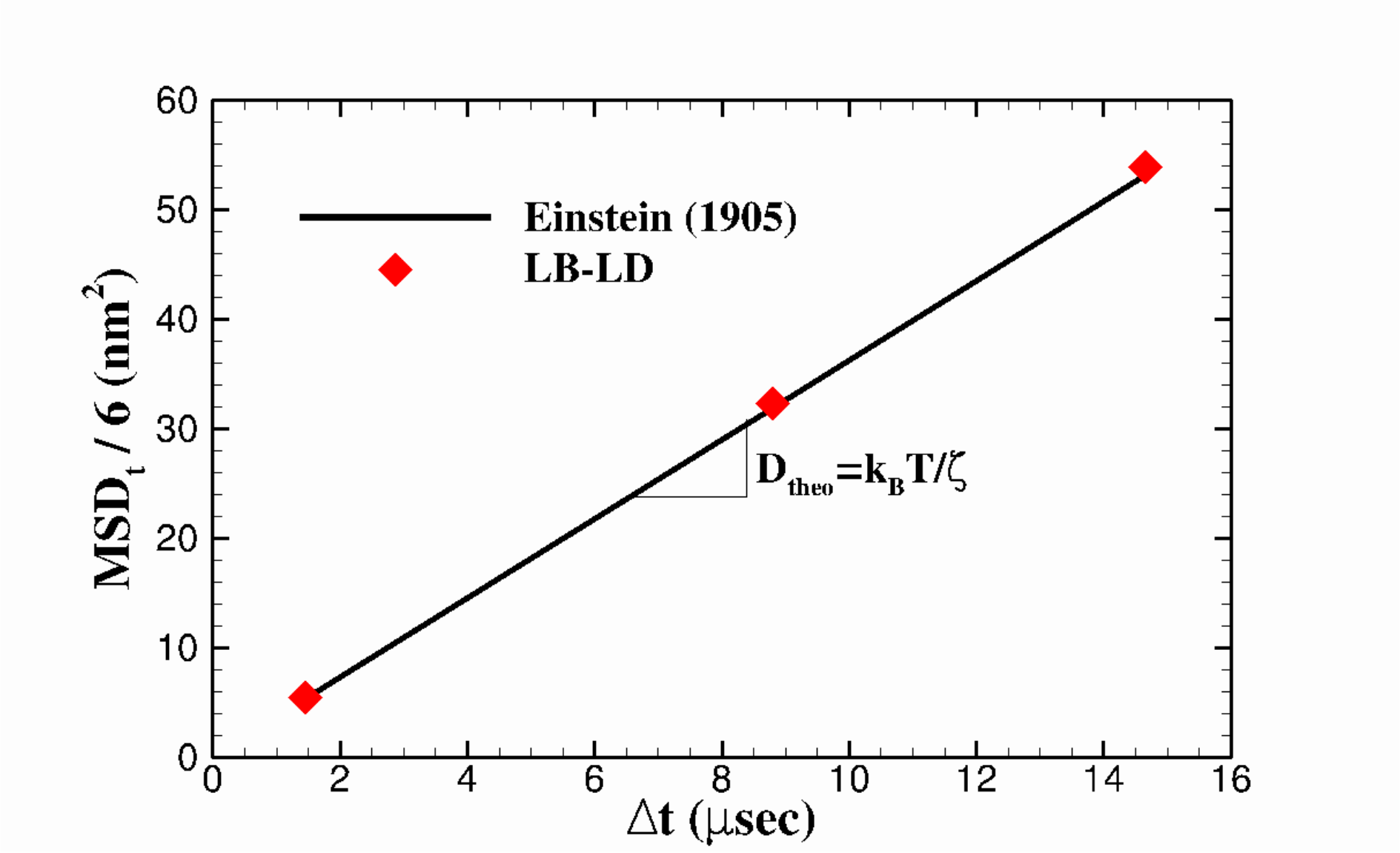

**Figure 8.** The scaled mean squared displacement, $MSD_t/6$, versus sampling time interval, $\Delta t$, for a $100\,nm$ particle self-diffusing in plasma at a temperature of $298\,K$. The diamond dot denotes the sampled long-time asymptotic $MSD_t/6$ for different $\Delta t$ with the solid line indicating the theoretical counterparts based on the Stokes-Einstein relation.

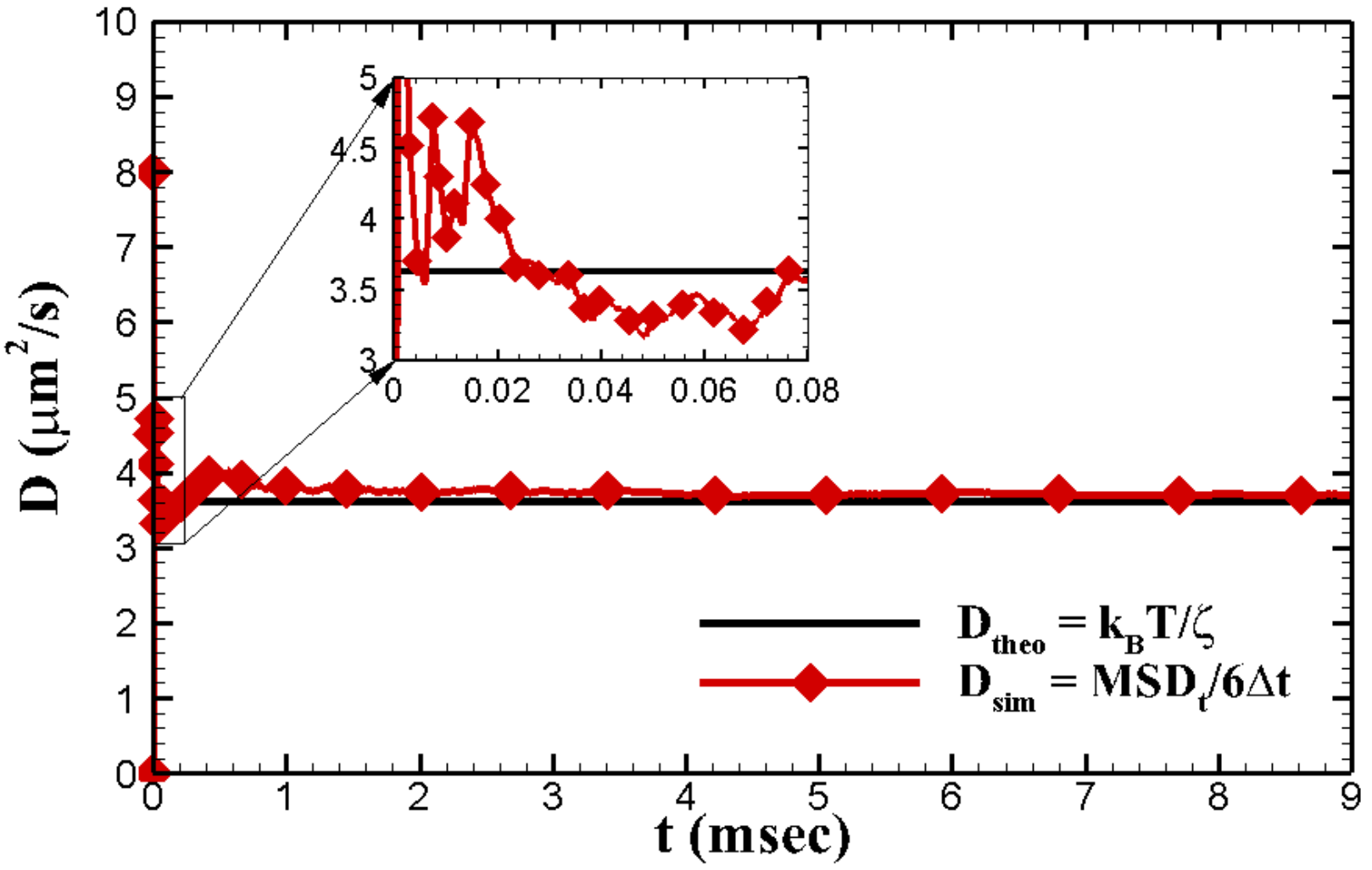

**Figure 9.** The temporal evolution of the particle diffusivity (lower), $D\,(\mu m^2/s)$, for a self-diffused $100\,nm$ particle in plasma at a temperature of $298\,K$. The sampling time interval, $\Delta t$, is fixed to be $14.7\,\mu sec$.

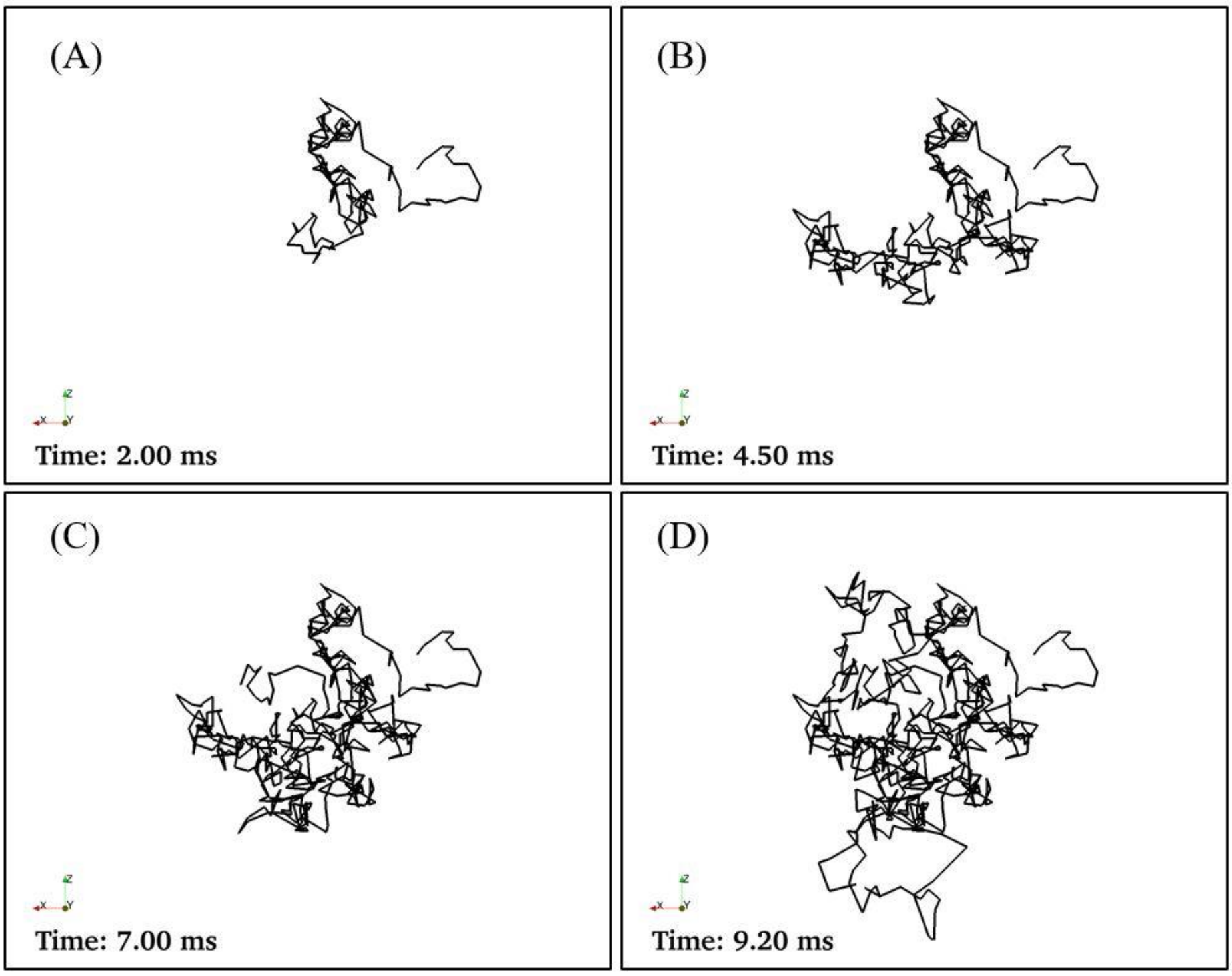

**Figure 10.** The temporal evolution of the particle trajectory (projected in the *x*-*z* plane) of a self-diffused $100\,nm$ particle in plasma at a temperature of $298\,K$. The trajectory sampling time interval, $\Delta t$, is $14.7\,\mu sec$.

Furthermore, a swarm of 10,000 particles of diameter $100\,nm$ is simulated to qualitatively demonstrate the effect of short-distance PP interaction on the particle diffusive behavior. Two initial NP packing concentrations are considered with the dilute case yielding a volume concentration of 0.5% and the dense case of 13%, denoted as (A) and (B), respectively. The particle MSD here is evaluated by further averaging the $MSD_t$ over the NP ensemble, denoted as $MSD_{t,e} = \overline{MSD_t}$. The sampling time interval also is fixed to be $\Delta t = 14.7\ \mu sec$. The instantaneous particle diffusivity is plotted against time in Figure 11. A much faster convergence to the analytical solution compared to the single particle case is observed for the dilute case due to considerable increase of samples. As time advances, the dilute case matches almost identically with the theoretical diffusivity, while the dense case gradually converges to a lower diffusivity. As shown in Figure 12 (B-2), the formation of particle clusters due to the near-field PP interactions (through LJ potentials) may explain the decrease of the particle diffusivity for the non-dilute case.

The analysis of self-diffusion for both single particle and particle swarm demonstrates that the current LB-LD hybrid approach correctly resolves the NP Brownian diffusive behavior. Unlike the previous studies [24, 27, 28] that include thermal fluctuation in both the particle phase and fluid phase, the current LB-LD method matches the Stokes-Einstein relation directly without empirically rescaling the friction coefficient.

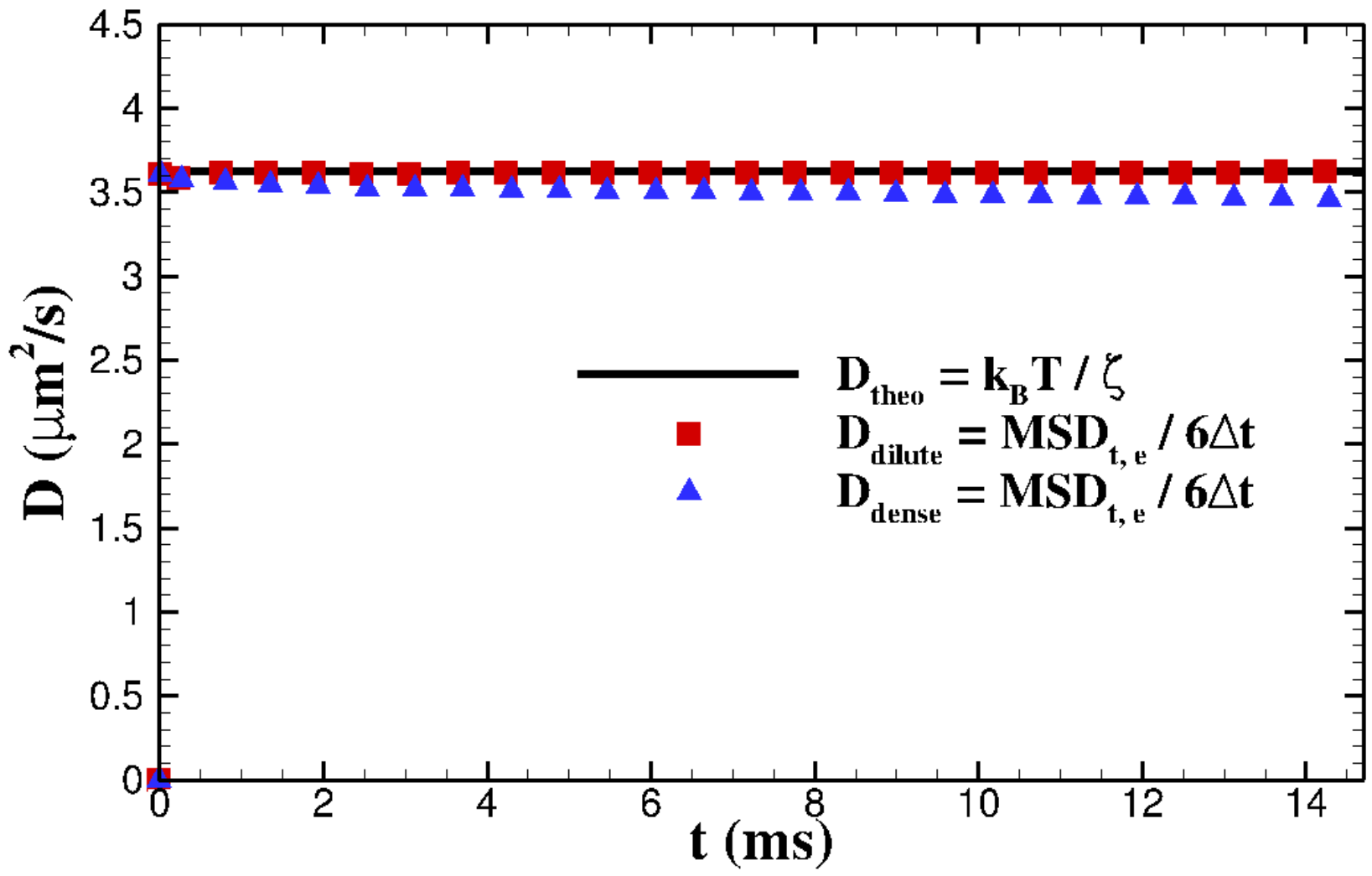

**Figure 11.** Temporal evolution of the particle diffusivity, $D\ (\mu m^2/s)$, for a $100\ nm$ particle swarm in plasma with initially dilute (A) and dense (B) distributions at a temperature of $298\ K$. The mean squared displacement, $MSD_{t,e}$, was calculated through both temporal and ensemble averaging with $\Delta t = 14.7\ \mu sec$.

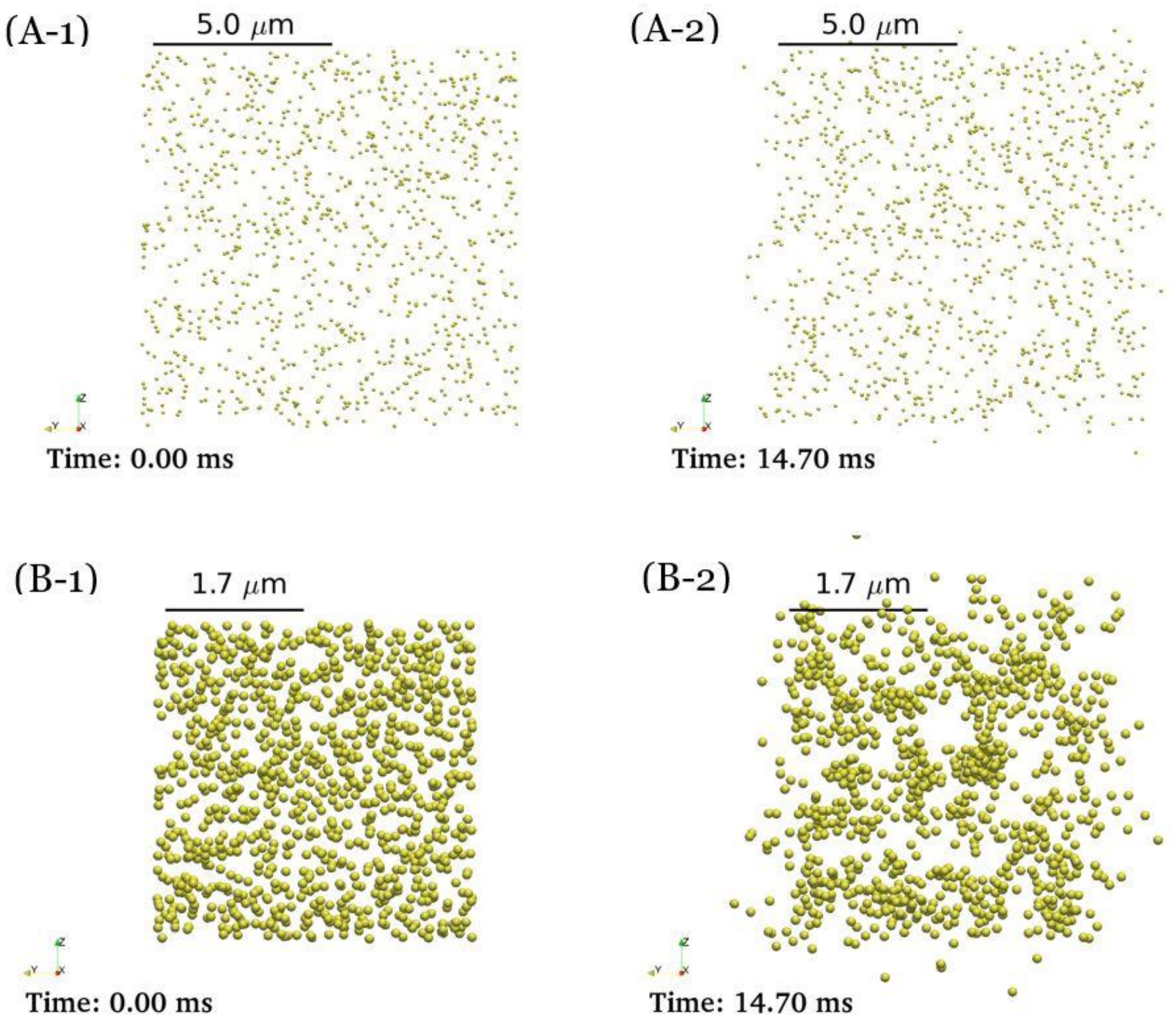

**Figure 12.** The beginning (left) and final (right) snapshots (projected in the *y-z* plane) of the diffusion process of a $100\ nm$ NP swarm with an initially dilute (A-1) or dense (B-1) distribution at a temperature of $298\ K$ in plasma.

## 4. APPLICATION

The multiscale computational approach is then applied to study the NP transport in a post-capillary vessel under physiological conditions. The RBC properties are assumed to match the physiological values, i.e., RBC diameter $d_p = 7.8\ \mu m$, RBC membrane shear moduli $G = 6.3 \times 10^{-6}\ J/m^2$, the cytoplasm to plasma viscosity ratio $\lambda = 5$, and the density of all the liquid phases $\rho = 1000\ kg/m^3$. The study below aims to shed light on how the particle Brownian motion affects the NP radial diffusion process.

### 4.1. *Simulation setup*

The relevant dimensionless parameters primarily include the vessel confinement ratio $d_v^*$, RBC hematocrit $\phi$, RBC capillary number $Ca_G$, and NP Péclet number $Pe$, as denoted in Figure 13. The confinement ratio is defined as the ratio between the vessel diameter $d_v$ and the RBC diameter $d_R$, i.e., $d_v^* = \frac{d_v}{d_R}$. The RBC hematocrit $\phi$ is defined as the volumetric percentage of RBCs in the blood vessel. The RBC capillary number reflects a competing effect between the fluid stress and membrane shear resistance, which is defined as

$$Ca_G = \frac{\mu \dot{\gamma}_w d_R}{2G}, \tag{34}$$

where $\dot{\gamma}_w$ is the shear rate at the vasculature wall. The RBC capillary number is a critical metric that largely affects the RBC dynamics and rheological properties of RBC suspensions [3, 7]. The NP Péclet number is defined as

$$Pe = \frac{3\pi\mu\dot{\gamma}_w d_P^3}{4k_B T}, \tag{35}$$

which quantifies the relative importance of the particle inertia over particle Brownian effect. Previous studies [3, 7, 11, 53] have shown that the RBC Reynolds number ($Re = \frac{\rho d_R^2 \dot{\gamma}_w}{\mu}$), if less than 1, exhibits negligible effects on the RBC dynamics as well as the particle margination propensity in micro-vessels. Therefore, current study intends to speed up the simulation by artificially scaling up the RBC Reynolds number to 1, while still matching $d_v^*$, $\phi$, $Ca_G$, and $Pe$.

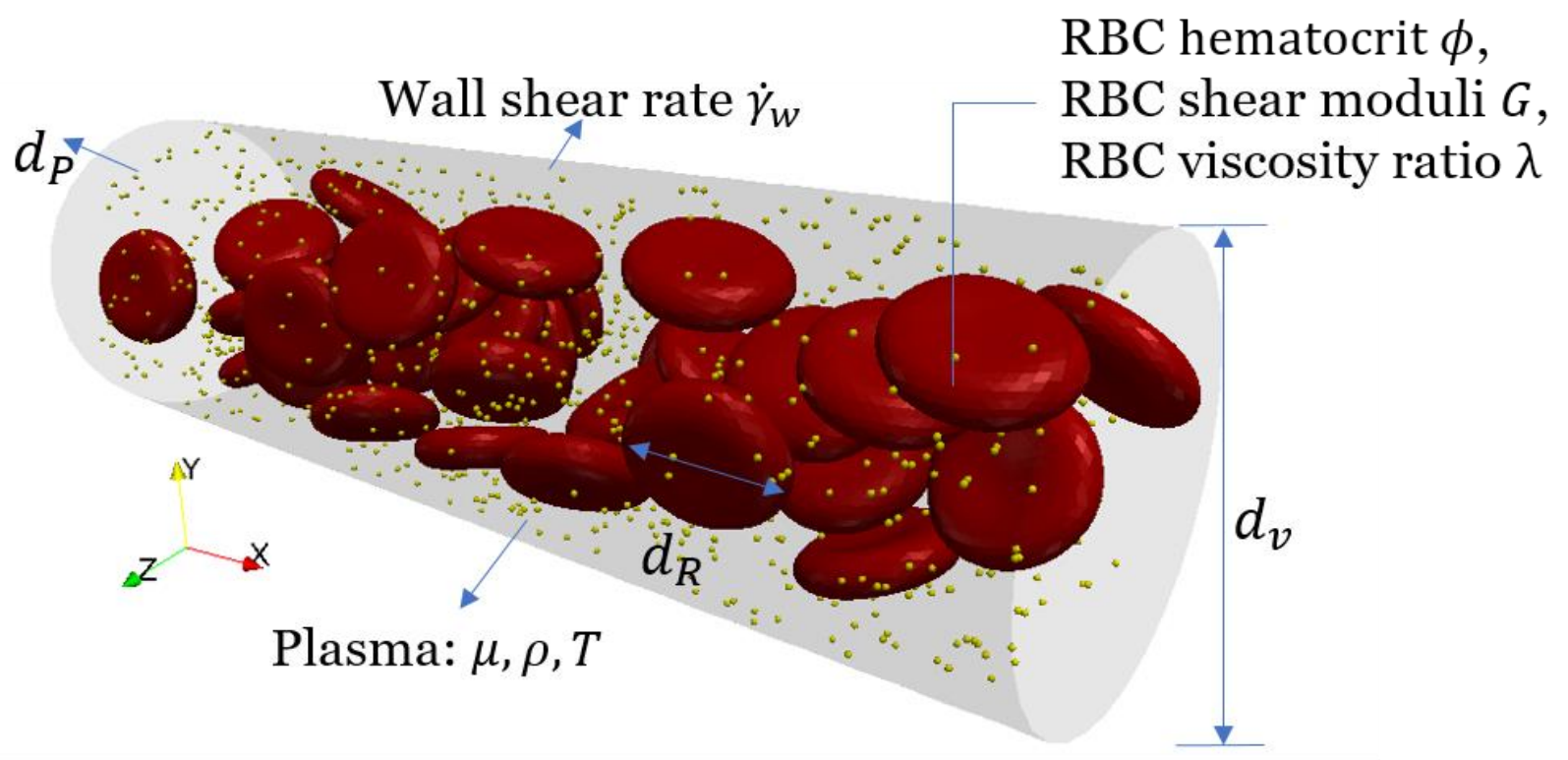

**Figure 13.** Instantaneous distribution of RBCs and $100\,nm$ particles in a $20\,\mu m$ diameter vessel with a RBC capillary number of $Ca_G = 0.37$ and a hematocrit $\phi = 20\%$; the relevant physical parameters are denoted accordingly. The tube axis is along the $x$ direction.

| Physiological ranges of relevant dimensional parameters | | Ranges of corresponding non-dimensional parameters | | Values of non-dimensional parameters selected |
|---|---|---|---|---|
| $d_v$ $(\mu m)$ | 8 ~ 30 | $d_v^*$ | 1.0 ~ 3.8 | 2.5 |
| $\phi$ | 10% ~ 26% | $\phi$ | 10% ~ 26% | 20% |
| $\dot{\gamma}_w$ $(s^{-1})$ | 50 ~ 1200 | $Ca_G$ | 0.04 ~ 0.90 | 0.37 |
| $d_p$ $(nm)$ | 50 ~ 500 | $Pe$ | 0.004 ~ 100.0 | 0.04, 0.34, 30.0, $\infty$ |

**Table 1.** Selection of relevant parameters for NP transport in a post-capillary vessel. The range of $\phi$ and $d_v$ is based on the experimental data reported by Fung [59]. The wall shear rate, $\dot{\gamma}_w$, is estimated based on the pressure drop in micro-vessels reported by Zweifach [60].

Table 1 tabulates the physiological ranges of the relevant dimensional and dimensionless parameters. The values of $\dot{\gamma}_w$ are estimated by $\dot{\gamma}_w = -\Delta P d_v / 4\mu$ based on the pressure drop, $\Delta P$, in micro-vessels reported by Zweifach [60]. The vessel diameter is fixed as $d_v = 20\ \mu m$, giving a confinement ratio of $d_v^* = 2.5$. The vessel wall shear rate is fixed to be $\dot{\gamma}_w = 460\ s^{-1}$, reflecting a Capillary number of $Ca_G = 0.37$. The hematocrit is chosen to be $20\%$. Three diameters of NPs, i.e. $50\ nm$, $100\ nm$, and $450\ nm$, are selected to vary the particle Brownian effect, which yields a $Pe$ of $0.04$, $0.34$ and $30.0$, respectively. $100\ nm$ particles with muted Brownian motion are simulated to solely demonstrate the RBC-enhanced diffusion effect, which is denoted as the $Pe = \infty$ case. A straight tube with a length-diameter ratio of 3 is used as a simplified vessel model; periodic boundary conditions are enforced on two ends of tube. Each tube section contains in total 34 RBCs and 1000 NPs. The simulation is started by concentrating the NP swarm on one end of the tube to mimic the initial condition when NPs are injected into the vessel, while RBCs are initially distributed in the NP-free section.

### 4.2. *Results and discussions*

First, the evolution of the normalized NP average radial locations, $\bar{r}/R_v$, are plotted for various $Pe$ in Figure 14, where $R_v$ is the radius of the vessel ($R_v = d_v/2$) and $\bar{r}$ is the NP average radial location. The average is carried out among all the NPs with number of particles $N = 1000$. Time is normalized by the wall shear rate as $\hat{t} = t\dot{\gamma}_w$. An increase of $\bar{r}/R_v$ is observed as $Pe$ decreases, showing the Brownian effect favors NPs to reach the vasculature wall. Comparing the $Pe = 0.34$ case with the $Pe = \infty$ case, about $20\%$ increase of the $\bar{r}/R_v$ is observed due to the presence of the NP Brownian effect.

The role of Brownian effect for NP margination can be further demonstrated by visualizing the NP distribution projected on the tube cross-section, as shown in Figure 15. When Brownian motion is present, as demonstrated in Figure 15 (B), NPs tend to be distributed over the entire cross-section of the tube. Considerable NPs are found closed to the vessel wall. In contrast, when Brownian effect is muted in Figure 15 (C), NPs tend to be more concentrated at the center region of the tube with only a minority of NPs being able to reach the vessel wall.

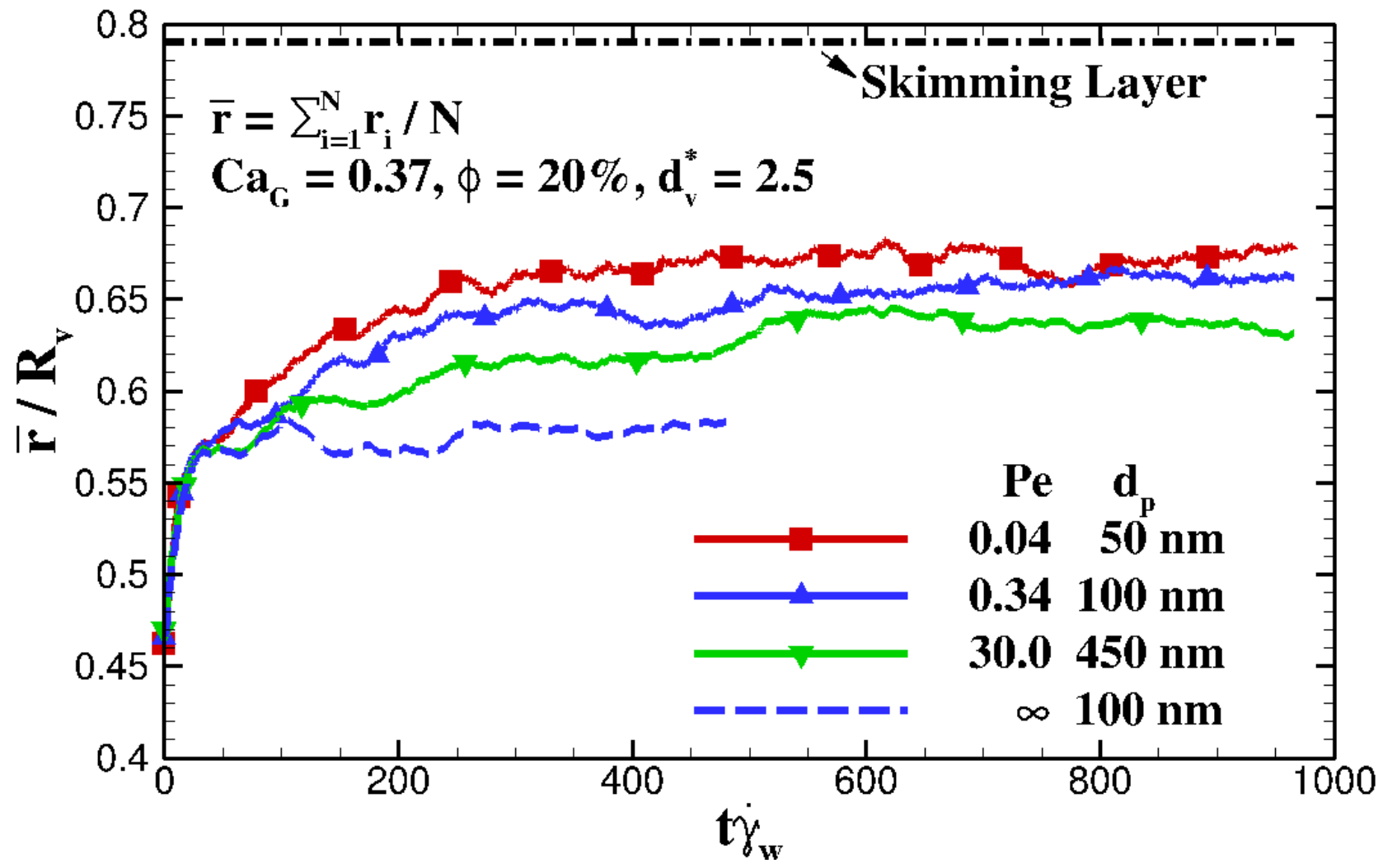

**Figure 14.** Temporal evolutions of normalized NP average radial locations for NPs sizing from $50\,nm$ to $450\,nm$ in a $20\,\mu m$ vessel with $\phi = 20\%$ and $Ca_G = 0.37$. The dash-dot line denotes the RBC skimming layer boundary location.

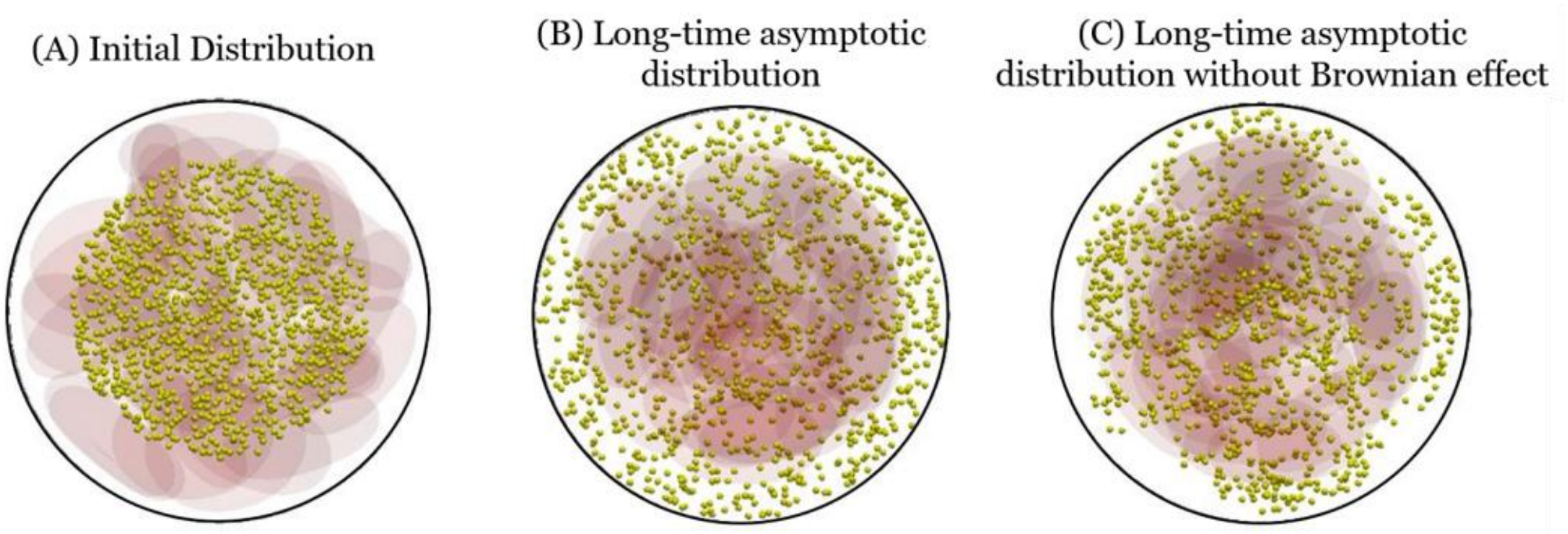

**Figure 15.** Instantaneous cross-sectional views (projected along $x$ direction) of $100\,nm$ particles (yellow dots) in a $20\,\mu m$ vessel with $\phi = 20\%$: (A) Initial distribution; (B) long-time asymptotic distribution with Brownian effect; (C) long-time asymptotic distribution without Brownian effect. RBCs are made semi-translucent to show all NPs.

The margination process of microscale particles (e.g., platelets) in blood vessels is primarily driven by the so-called RBC-enhanced diffusivity, which originates from the rigid particles interacting with deformable RBCs as well as the sink-like effect of the RBC depleted zone [11, 14, 17]. Given Brownian motion plays a significant role in the NP radial diffusion process as demonstrated above, the particle total radial diffusivity $D_{rr}$ can be reasonably assumed to be decomposed into the Brownian diffusivity $D_{rr}^{B}$ and RBC-enhanced diffusivity $D_{rr}^{R}$, i.e., $D_{rr} = D_{rr}^{B} + D_{rr}^{R}$. By measuring the particle MSD along radial direction, the averaged total diffusivity of NPs can be evaluated as

$$D_{rr} = \left. \frac{\langle \left[ r_p(t + \Delta t) - r_p(t) \right]^2 \rangle_{e,t}}{2\Delta t} \right|_{\Delta t \gg \tau_r}, \tag{36}$$

where the sampling time interval is chosen to be the same as that used to verify the particle diffusive behavior in Section 3.2.

The temporal evolutions of the average NP radial diffusivity for different $Pe$ are shown in Figure 16. The dash-dot lines indicate the theoretical Brownian diffusivity evaluated by Einstein's relation for each $Pe$. During the initial stage, all cases exhibit a sudden increase of the total diffusivity, reflecting the transient NP-RBC mixing process. Within $\sim 2\ s$, the NP diffusion process approaches the long-time asymptotic value.

In the long-time asymptotic regime, the $Pe = 0.04$ case produces an total diffusivity, $D_{rr}$, almost identical to the theoretical Brownian diffusivity, $D_{rr}^{B}$; for the case with $Pe = 0.34$, $D_{rr}$ converges to a value slightly greater than $D_{rr}^{B}$; the $Pe = 30$ case produces a $D_{rr}$ that is $\sim 1.6$ folds of the $D_{rr}^{B}$. The $d_p = 100\ nm$ case with muted Brownian effect ($Pe \sim \infty$) produces a much lower diffusivity than the same case with Brownian effect ($Pe =$ 0.34). The diffusivity difference is about the same as the Brownian diffusivity of the $d_p = 100\,nm$ case, implying $D_{rr} = D^B_{rr} + D^R_{rr}$ is an acceptable assumption here. By quantifying the diffusivity ratio $D^B_{rr}/D^R_{rr} = D^B_{rr}/(D_{rr} - D^B_{rr})$, it can be shown that both cases with $Pe < 1$ satisfy $D^B_{rr}/D^R_{rr} \gg 1$, indicating that Brownian diffusion dominates the radial diffusion process of NPs of diameter $d_p \leq 100\,nm$. As particle diameter increases to $\sim 500\,nm$, this diffusivity ratio gets close to unity, $D^B_{rr}/D^R_{rr} \sim 1$, suggesting that both RBC-enhanced diffusion and Brownian diffusion contribute comparably to NP radial diffusion. As particle diameter further increases to microscale, Brownian effect becomes negligible, RBC-enhanced diffusion starts to take over the particle migration process ($D^B_{rr}/D^R_{rr} \ll 1$) [11, 14].

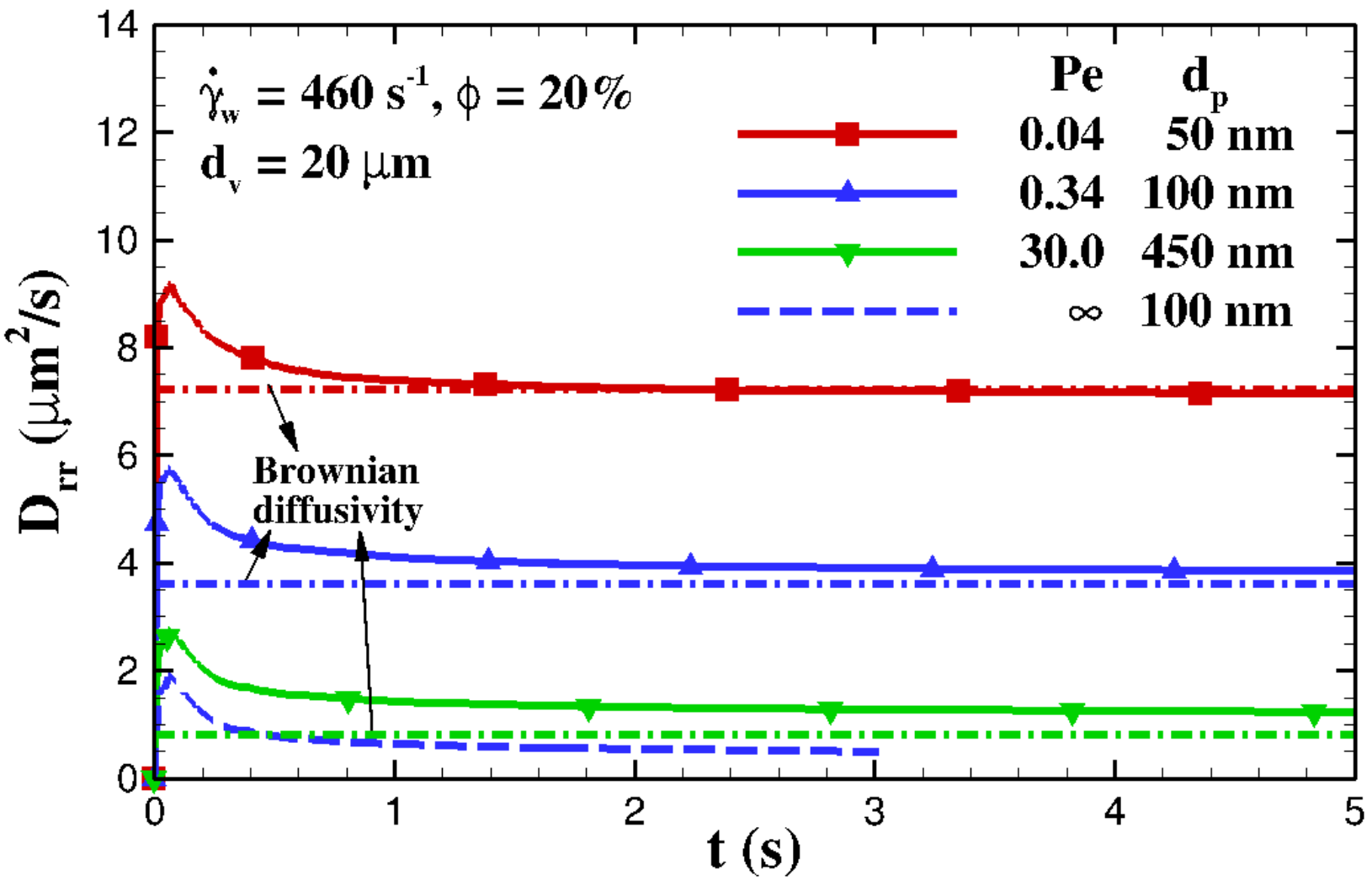

**Figure 16.** Total NP radial diffusivity, $D_{rr}$, plotted against time for various Péclet numbers with corresponding theoretical Brownian diffusivities (dash-dot lines) given as comparisons. Simulations are performed with $d_v = 20\,\mu m$, $\dot{\gamma}_w = 460\,s^{-1}$ and $\phi = 20\%$.

## 5. CONCLUSIONS

A comprehensive multiscale approach that couples the LB, LD and SL methods has been developed and verified for simulating NP transport in cellular blood flow with high efficiency. Both the particle-fluid coupled dynamic response and the long-time particle Brownian diffusive behavior have been demonstrated to match well with classical results. This multiscale computational approach is expected to be an accessible *in-silico* device to support the development of nanotherapeutic systems. To name a few of its applications, the LB-LD-SL approach can be used to study the NP transport mechanisms in cellular blood flow, to obtain rheological and transport properties for phenomenological models, and to narrow down the parameter space to aid experimental design.

In the study of NP diffusion in a post-capillary vessel, the NP radial diffusive mechanisms are studied and decomposed quantitatively. The relative importance of RBC-enhanced diffusion and Brownian diffusion is discussed in detail. Brownian diffusion and RBC-enhanced diffusion are both shown to be comparably important drivers, i.e., $D_{rr}^{B}/D_{rr}^{R} \sim 1$, for particles of diameter $\sim 500\, nm$ ($Pe \sim 30$); however, when particle diameter is smaller than $100\, nm$ ($Pe < 1$), Brownian diffusion becomes the major mechanism that dominates the NP radial diffusion process, producing $D_{rr}^{B}/D_{rr}^{R} \gg 1$. Overall, Brownian motion is shown to be favorable for the margination of NPs in post-capillary vessels. This study together with the studies of micro-sized particle margination (where $D_{rr}^{B}/D_{rr}^{R} \ll 1$) in cellular blood flow [11, 14, 17] presents a multiscale understanding on the diffusive mechanisms of nano- and micro-sized particle in cellular blood flow.

Given the variety of the cellular environment and NP characteristics, other factors such as vasculature confinement, hematocrit, RBC deformability, and NP properties (e.g., shape, surface decorations) need to be further taken into consideration to eventually obtain a complete parametric relation of the NP radial diffusivity in micro-vessels.

## ACKNOWLEDGEMENTS

Z. Liu is funded by Sandia National Laboratories under grant 2506X36. The computational resource is granted by the Extreme Science and Engineering Discovery Environment (XSEDE) of National Science Foundation under grant TG-CT100012.

Sandia National Laboratories is a multimission laboratory managed and operated by National Technology and Engineering Solutions of Sandia LLC, a wholly owned subsidiary of Honeywell International Inc. for the U.S. Department of Energy's National Nuclear Security Administration under contract DE-NA0003525.